# Thickness-dependent degradation and optical access in epitaxial 2H-$MoTe_2$ protected by metallic capping layers

Wojciech Ryś[1], Iaroslav Lutsyk[1], Michał Piskorski[1], Maxime Le Ster[1], Maciej Rogala[1], Paweł Dąbrowski[1], Paweł Krukowski[1], Katarzyna Ranoszek-Soliwoda[2], Jarosław Grobelny[2], Zuzanna Ogorzałek-Sory[3], Bartłomiej Seredyński[4], Wojciech Pacuski[3], Janusz Sadowski[5,6], Marta Gryglas-Borysiewicz[3], Karol Szałowski[1], Paweł J. Kowalczyk[1*]

[1] University of Lodz, Faculty of Physics and Applied Informatics, Pomorska 149/153, 90-236 Łódź, Poland

[2] University of Lodz, Faculty of Chemistry, Pomorska 163, 90-236 Lodz, Poland

[3] University of Warsaw, Faculty of Physics, Pasteura 5, 02-093 Warsaw, Poland

[4] Military University of Technology, Institute of Applied Physics, Kaliskiego 2, 00-908 Warsaw, Poland

[5] Polish Academy of Sciences, Institute of Physics, Aleja Lotnikow 32/46, Warsaw, Poland

[6] Ensemble3 Centre of Excellence, Wolczynska 133, Warsaw, Poland

## Abstract

We investigate degradation and surface protection of epitaxial 2H-$MoTe_2$ films grown by molecular beam epitaxy on GaAs(111)B substrates. Using X-ray photoelectron spectroscopy (XPS), scanning tunneling microscopy, atomic force microscopy (AFM), Kelvin probe microscopy (KPM), Raman spectroscopy, and density functional theory (DFT), we examine the structural, chemical, and electronic evolution of $MoTe_2$ protected by Co and Ni capping layers. XPS shows that the metallic caps effectively suppress oxidation during short-term air transfer, while revealing a pronounced Te-rich near-surface composition. With time, the caps become increasingly difficult to remove, suggesting gradual interfacial bonding promoted by excess tellurium and defect-rich $MoTe_2$ interfaces. AFM and KPM reveal pronounced thickness-dependent ageing, with ultrathin regions showing markedly different contact-potential evolution from thicker films. DFT calculations support the sensitivity of work function and density of states to thickness and surface chemistry. Raman measurements through approximately 20 nm thick metallic caps demonstrate partial optical access to the protected material. Additional AFM and Raman observations suggest local formation of Te-rich nanostructures under laser illumination or near mechanically damaged regions. These results provide practical guidelines for protecting, transferring, delaminating, and optically characterizing air-sensitive $MoTe_2$ and related van der Waals materials.

## 1. Introduction

Two-dimensional (2D) transition metal dichalcogenides (TMDCs) such as 2H-$MoTe_2$ have attracted significant interest due to their layer-dependent electronic [1–6] and optical properties [1–3,6,7], making them promising candidates for next-generation nanoelectronic [1–3,8,9] and optoelectronic devices [1–3,7,10–13]. $MoTe_2$ is particularly notable for its polymorphic behavior – 2H and 1T’ phases often coexist at room temperature [14–19] (1T' formation energy per formula unit is several meV higher than that of 2H [20]; see Fig. 1a and Fig. 1b for ball and stick models). $MoTe_2$ in the 2H phase is a direct bandgap semiconductor, with a gap width of 1.1 eV in its monolayer phase [14,21], decreasing with thickness [7,22,23]. On the other hand, the 1T' phase shows metallic properties at RT but undergoes a phase transition below 250 K, forming the orthorhombic Td phase, becoming a type-II Weyl semimetal [15,19]. Recently, modulation of the superconducting gap magnitude, coherence strength, and subgap states were observed in monolayer 1T'-$MoTe_2$, revealing unconventional superconducting behavior that deviates from the standard Bardeen-Cooper-Schrieffer description [24]. It was also shown that transitions between semiconducting and metallic phases are possible under external stimuli [17,18,24,25], which makes $MoTe_2$ a prospective material in future applications [16,19,27,28], for instance, as a low-Schottky-barrier contact in $MoS_2$-based electronics [29].

However, the practical use of $MoTe_2$ is subject to several limitations. It has recently been shown that $MoTe_2$ can be modified by laser irradiation, resulting in the appearance of metallic Te signatures in Raman spectroscopy [30,31]. In earlier studies, similar Te-like Raman features were often attributed to the formation of the 1T' phase of $MoTe_2$ [16,32–43] which may require reconsideration in light of recent findings [30,31]. Typically, the presence of Te and its nanostructures is associated with a mode in the 120–128 $cm^{-1}$ range, accompanied by a feature near 140 $cm^{-1}$ and a weaker signal around 103 $cm^{-1}$ [30,31,44–46]. The position of the main mode (i.e., 140 $cm^{-1}$) depends strongly on morphology and may appear even near 115 $cm^{-1}$ [47]. It was also reported that as the size of Te nanowires decreases, the modes shift toward higher wavenumbers, with the main feature located above 130 $cm^{-1}$ [44].

Another key limitation for the practical use of $MoTe_2$ is its poor environmental stability [48–51]. Even short-term exposure to air can lead to rapid surface degradation, particularly in the limit of ultrathin layers [50–54]. This issue is especially critical for epitaxial films grown under ultra-high vacuum (UHV) conditions, where transferring samples between growth and characterization systems often introduces surface contamination unless protective measures are undertaken. To address this issue, several capping strategies have been explored [48,51,55–58]. Among them, hexagonal boron nitride (hBN) is considered one of the most promising materials due to its electronically insulating nature, transparency, and high dielectric constant [17,55,59–61], which enable a wide range of optoelectronic measurements on air-sensitive materials [62–65]. However, hBN flakes are typically transferred mechanically in air or within a glovebox – methods that can induce initial surface degradation and are difficult to handle. Graphene, the other prospective 2D material, has also been shown to act as an effective antioxidant and transparent capping layer [56,58]. Nevertheless, structural defects in graphene can lead to localized degradation [66], and its semi-metallic nature may interfere with electrical or optoelectronic measurements [67]. While these encapsulation techniques offer significant promise, they remain challenging to implement for samples grown in UHV. In such cases, thin metallic capping layers are often deposited *in situ* directly onto the freshly grown material [48,51,57]. If sufficiently thin, these layers may serve a dual purpose: providing temporary protection and enabling further characterization without the need for sample surface exposure to air [51].

In this work, we investigate the degradation behavior of epitaxial 2H-$MoTe_2$ films grown on GaAs(111)B (As terminated) substrates [48,51], with a particular focus on thickness-dependent oxidation and surface protection using thin metallic capping layers of Co and Ni. Using a combination of surface-sensitive techniques, including X-ray photoelectron spectroscopy (XPS), atomic force microscopy (AFM), scanning tunneling microscopy (STM), Raman spectroscopy, and Kelvin probe microscopy (KPM), we reveal a strong thickness dependence of both the degradation rate and electronic properties (i.e., contact potential difference – CPD and inversely proportional work function

– WF) of $MoTe_2$. STM measurements reveal domain-boundary reconstructions reminiscent of previously reported inversion domain defects [68]. However, instead of the characteristic triangular features, we observe distorted asterisk-like patterns that we attribute to the presence of locally twisted domains and Te-deficient regions. In contrast, XPS analysis indicates an overall tellurium enrichment at the surface, suggesting that while the STM-probed areas are Te-deficient, the film as a whole contains Te accumulated near the surface. Raman spectroscopy and AFM further suggest that Te-rich regions may develop locally, especially on old samples after laser exposure (leading to local heating [31,69]) or mechanical damage. Moreover, we demonstrate that 2H-$MoTe_2$ degrades rapidly under ambient conditions. However, thin Co or Ni capping layers provide effective short-term protection, allowing safe transfer of samples between laboratories. These layers can initially be removed by mechanical exfoliation, revealing oxide-free $MoTe_2$ surfaces. After several days, however, the metallic layers appear to develop stronger interfacial bonding with the underlying material, preventing successful delamination. Further, we show that the oxidation behavior of $MoTe_2$ is thickness-dependent, as supported by our DFT calculations. Finally, we demonstrate that sufficiently thin metallic capping layers can simultaneously protect the material against oxidation and serve as semitransparent barriers enabling optical characterization of the encapsulated $MoTe_2$ films without removal of the overlayer.

Altogether, this work provides new insight into the degradation pathways of 2H-$MoTe_2$, the dual role of metallic capping layers in both protecting and altering the surface, and the possible local emergence of Te-rich nanostructures, offering practical considerations for handling and characterizing air-sensitive 2D materials.

## 2. Results

Metal (Co, Ni) capped $MoTe_2$ films were grown by MBE on GaAs(111)B substrates (samples with nominal thickness of both 10-layer (10 L) or 4-layer (4 L) of 2H-$MoTe_2$ were used) [51]. The experimental capping/decapping procedure is depicted in Fig. 1c. On the as-grown $MoTe_2$ layers thr protective Co or Ni metal capping layers are deposited in-situ. After the transfer of the sample through air from the growth chamber to the characterization laboratory, the metallic layer is removed with an adhesive tape. Depending on the experiments, the exfoliation occurred either in a vacuum or an argon-filled glove box. This procedure allows for measurements of the bare $MoTe_2$ surface, while the resulting adhesive tape offers access to the $MoTe_2$, which was peeled off. In particular, samples for STM, low energy electron diffraction (LEED) and XPS investigations were exfoliated in UHV, next exposed to air for AFM measurements and again loaded into UHV for further XPS studies. Afterwards, these samples were also studied using Raman spectroscopy in order to carry on measurements in the same spots over the surface. In turn, for measurements on adhesive tape the sample was exfoliated in glove box and studied immediately after preparation.

### 2.1. Crystallographic identification using STM and LEED

The 4 L thick samples were investigated using STM after exfoliation in UHV. A representative atomic-resolution image is shown in Fig. 1d. The surface exhibits a complex morphology with brighter regions composed of nanorod-like features (a few nanometers in length) arranged around local centers to form characteristic asterisk-like patterns (see white dashed ovals in Fig. 1d). In many cases, these patterns gradually evolve into triangular networks (see upper inset in Fig. 1d), previously identified as inversion domain boundary defects [68] associated with twin domains rotated by 60° [70–72]. Such boundaries are known to be Te-poor, as evidenced by thermal reduction experiments that increase the density of triangular superstructures while reducing their size [70].

Closer inspection of both the asterisk-like structures and the triangular boundaries reveals small angular distortions: nanorods and boundaries are rotated by several degrees relative to the ideal lattice orientation (see blue and green arrows in Fig. 1d and the upper inset). These rotations suggest a domain-like growth mode in which neighboring domains are misaligned by a few degrees beyond the typical 60° rotation reported previously [70–72]. The resulting strain hinders the closure of triangular patterns, giving rise to the observed open, asterisk-like structures. We therefore interpret these nanorods as domain boundaries between slightly misoriented regions, where incomplete triangles manifest as extended radial motifs.

While the majority of the surface is covered by such domain boundaries, we also identify regions with well-ordered triangular atomic patterns (see lower inset in Fig. 1d), characteristic of stoichiometric 2H-$MoTe_2$ [70]. The overall STM observations are consistent with LEED patterns (inset in Fig. 1d), which show triangular symmetry and confirm that the films are in the 2H phase [51]. Overall, our measurements suggest that the growth is not entirely homogenous and results in a mixture of stoichiometric regions, Te-poor boundaries, and asterisk-like reconstructions arising from slight misorientation between neighboring domains.

## 2.2. XPS measurements

XPS measurements were performed to demonstrate the protective effect of Co capping layer against surface oxidation and to characterize the air-exposure-induced degradation of 2H-$MoTe_2$. First, a clean surface of 10 L thick 2H-$MoTe_2$ was investigated after removing a capping layer in UHV conditions – see upper spectrum denoted 01 in Fig. 1e,f. The spectra are characterized by two doublets located at 228.0, 231.1 eV and 573.0, 583.3 eV, which are related to 3d5/2, 3d3/2 of Te and Mo, respectively [32,73]. The presence of uniform Mo and Te core level lines in our spectra confirms that the surface was effectively protected by the capping layer and that no oxides were formed. Interestingly, the Te:Mo ratio calculated from the observed peak intensities is close to 4, indicating an excess of Te in the surface region (see further comments in the Discussion section).

A significant decrease in the intensity of Mo 3d and Te 3d peaks (characteristic of stoichiometric $MoTe_2$), together with the appearance of new doublets at 232.4, 235.6 eV and 576.3, 586.8 eV, is observed after exposing the sample to air for 7 hours (see spectrum 02 in Fig. 1e,f). These new features correspond to the oxidation of the 2H-$MoTe_2$ surface, leading to the formation of O-Te and O-Mo bonds. Quantitative analysis shows a pronounced reduction of metallic Mo in favor of oxidized Mo, with an estimated Mo:O-Mo ratio of 0.59, indicating that nearly two-thirds of Mo atoms are oxidized within the probed near-surface region. In contrast, metallic Te remains dominant over tellurium oxide, with a ratio of about 1.85, suggesting that less than half of the Te atoms are oxidized. Notably, the Te:Mo ratio for the unoxidized components increases to approximately 5, further highlighting the preferential oxidation of Mo.

Interestingly, the oxygen content in the surface layer barely changes after 2.5 years in air (see spectrum 03 in Fig. 1e,f). This indicates that the oxidation is a relatively rapid process, taking place over the course of the first hours following the initial exposure of the sample to air. The oxide appears to passivate the surface, considerably limiting further oxidation. The ratio of Te:Mo content reduces back to 4.1, Mo:O-Mo does not change, while Te:O-Te is reduced to 1.75. These observations indicate that oxidation predominantly affects the excess tellurium atoms.

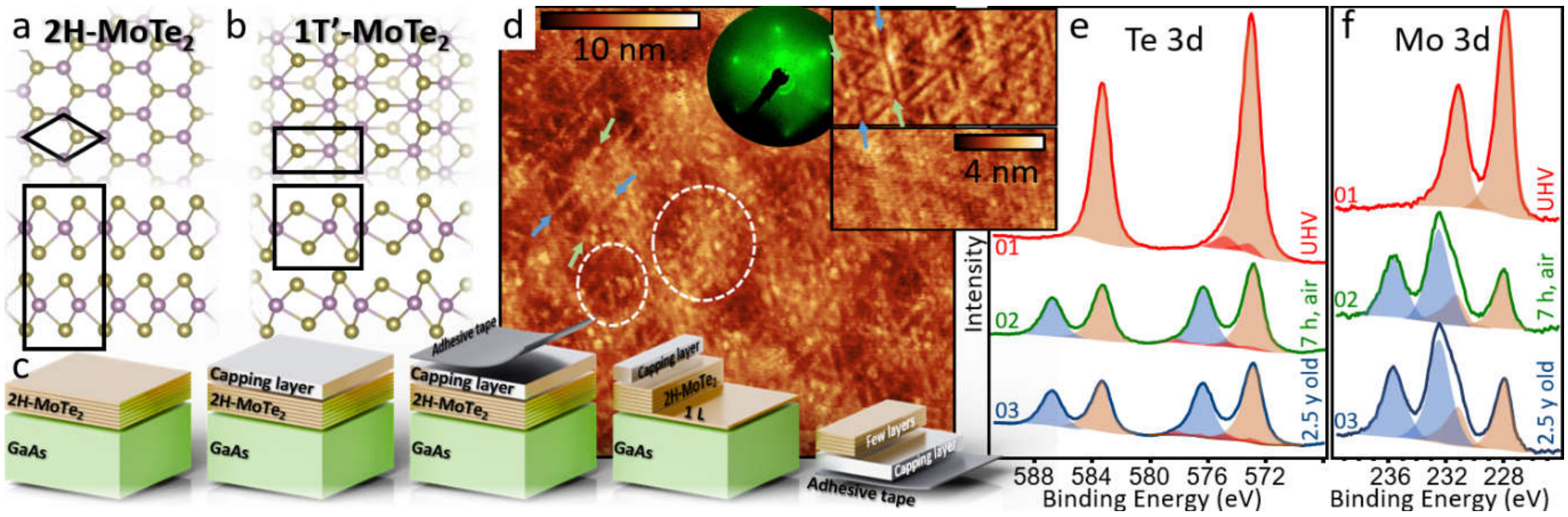

*Fig. 1. The ball and stick models of (a) 2H-$MoTe_2$ and (b) 1T'-$MoTe_2$ phases. (c) A schematic representation of the samples' preparation process. (d) STM images of the $MoTe_2$ surface after Ni layer removal (main: I=0.8 nA, U=20 mV; top right: I=1.0 nA, U=500 mV; bottom right: I=1.0 nA, U=300 mV). Arrows show twisted grain boundaries, dashed ovals show asterisk-like patterns described in the main text. The LEED pattern acquired at 65 eV is shown in the inset. The XPS spectra for (e) Te 3d and (f) Mo 3d subshells of 2H-$MoTe_2$ sample 01: exfoliated in UHV, 02: exposed to air for 7 hours, and 03: 2.5 years old.*

## 2.3. Morphological Characterization

### *2.3.1. Optical microscopy*

A typical morphology observed using optical microscopy (OM) of samples after peeling off the capping layer is shown in Fig. 2a. Contrast in OM is related to the thickness of the examined structures, namely, darker regions in OM correspond to thinner structures. This is confirmed by OM contrast simulations shown as a scale bar on the right in Fig. 2a, in which the expected change of a color as a function of $MoTe_2$ thickness is shown [31]. The complex refractive indices used in the simulations across this work were taken from Ref. [55] and for 532 nm light are as follows: Ni (1.0660, 4.9183), Co (0.6356, 4.5414), $MoTe_2$ (4.9513, 2.6100), and GaAs (4.0668, 0.3487), where the first and second values in parentheses denote *n* (refractive index) and *k* (extinction coefficient), respectively.

Inspection of the OM image shown in Fig. 2a suggests that the removal of the Co capping layer leads to the formation of vast areas of tens of square micrometers with homogeneous thicknesses. At first glance, three areas can be distinguished – the brightest are associated with the fragments of the cobalt capping layer (labeled Co cap in Fig. 2a), which initially covered the entire sample before exfoliation. The darkest ones are related to 1 L (regions R1 and R2 in Fig. 2) and intermediate ones, to 9 L $MoTe_2$ (see R3 and R4 regions in Fig. 2; for thickness assignment, see discussion below).

### *2.3.2. Atomic force microscopy*

A more detailed surface study is carried out using AFM microscopy, with typical topography shown in Fig. 2b recorded in the area indicated by a black square in Fig. 2a. It is characterized by the presence of 50 nm tall "pyramids" in the GaAs(111)B surface, which are not resolved in OM images. In the central part of the image, the 17.2 ± 1.5 nm tall Co island is observed, a remnant of the capping layer (labeled Co cap in Fig. 2b). The height of R3 and R4 regions measured with respect to the base layer is 5.7 ± 1.0 nm (i.e., 8 L [74–76]). Occasionally, regions thinner than this base (by approx. 1 L) are observed, which we identify as distorted $MoTe_2$ exhibiting Te-related Raman features (see region marked distorted/Te-rich in Fig. 2b and cross section c-2 in SI Fig. S1). Overall, the samples are characterized by the coexistence of 1 L and 9 L thick regions (formed on top of distorted $MoTe_2$) which indicates that 9 L and 1 L are removed upon capping layer removal.

### *2.3.3. Contact potential difference*

We now discuss the contact potential difference (CPD) measured using KPM to detect the possible inhomogeneities in the 2.5-year-old, i.e., heavily oxidized $MoTe_2$ sample (see spectrum 03 in the XPS data shown in Fig. 1e,f). Note, air-exposed samples are covered by a layer of water and atmospheric contaminants [77], which in general leads to a decrease in measured WF [77–80]. As a consequence, observations related to CPD and WF discussed here are qualitative.

The CPD image recorded together with the AFM topography (Fig. 2b) is shown in Fig. 2c. It is characterized by the presence of areas of homogeneous contrast, the geometric shapes of which indicate a direct relationship between the CPD and the thickness of the investigated region. These data clearly show that $CPD_{1L}$>$CPD_{9L}$ (where subscript denotes thickness). Keeping in mind that, in first approximation, CPD is inversely proportional to the WF, we conclude that the work function for oxidized $MoTe_2$ is thickness dependent (see also DFT calculations discussed below). Moreover, it is clearly visible that the CPD is insensitive to the presence of GaAs pyramids. This indicates the homogeneity of the electronic properties of the $MoTe_2$ layers, regardless of their local morphology (note that the spatial resolution of KPM is at the level of dozens of nanometers and consequently local perturbations of CPD are beyond the detection limit in our setup). Finally, we point out that the CPD measured on distorted $MoTe_2$ is lower than on base 1 L thick regions (see Fig. 2c).

To better understand the oxidation process of $MoTe_2$, we compare the results discussed above with the data recorded on the same sample but when it was fresh – exposed to air for the short time (see Fig. 2d; measurements conducted in the region indicated by the yellow square in Fig. 2c). This image clearly shows that in the initial oxidation phase, the CPDs for 1 and 9 L are very similar (difference at the level of measurement error of 30 mV). Clearly, after prolonged exposure, the CPD of the 1 L region increases noticeably, while only minor changes are observed for the 9 L region and the Co capping layer. This suggests that the observed CPD evolution may arise from thickness-dependent oxidation and/or from different interfacial environments underneath 1 L and 9 L regions, as discussed further in the text.

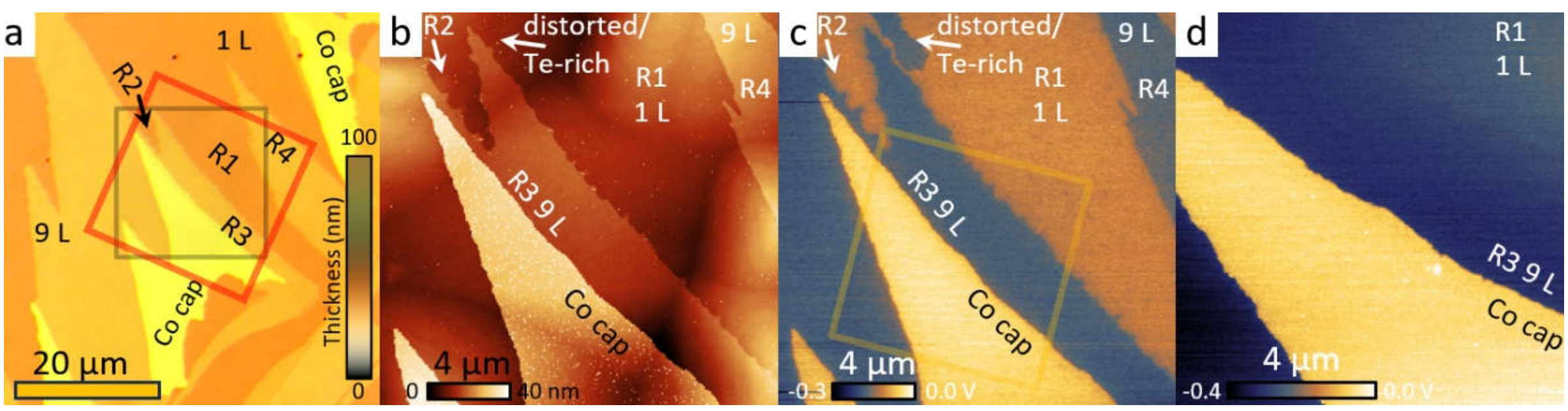

*Fig. 2. (a) Optical microscopy image showing the morphology of $MoTe_2$ grown on GaAs(111). (b) AFM recorded on 2.5 yr old sample in the region indicated using a black square in (a). (c) KPM image recorded together with topographic data shown in (b). (d) KPM image recorded on few hours old sample in the region indicated using yellow square in (c).*

## 2.4. Raman investigations of $MoTe_2$

In this section, we describe the Raman spectroscopy results recorded on GaAs substrates, $MoTe_2$ films grown on GaAs, exfoliated $MoTe_2$ on adhesive tape, and reference spectra of 2H and 1T' phases of bulk $MoTe_2$ crystals (see Fig. 3a and SI Fig. S2a for individual spectra). Inspection of the data in Fig. 3a shows that spectra recorded on $MoTe_2$/GaAs are dominated by GaAs modes located at Raman shifts of 267 (TO), and 291 $cm^{-1}$ (LO) and to a lesser extent by modes at 124 and 159, 333, 505, 533, 573 $cm^{-1}$ (see spectra 01-03 and the gray areas in Fig. 3c and SI Fig. S2a, in particular 02 was recorded on the substrate prepared in a similar way as for the growth i.e., by annealing in UHV while 01 was recorded on the scratched substrate and shows least amount of modes; in turn 03 is recorded on heavily oxidized substrate). These modes characteristic of GaAs are consistent with previous studies, although earlier reports focused primarily on the TO and LO modes [81–83]. Raman spectroscopy for $MoTe_2$ is discussed below.

### *2.4.1. Raman mapping*

Raman spectroscopy mapping enables spatial visualization of the Raman mode intensity variations across the sample surface. In Fig. 3b, a map for the GaAs TO mode recorded at the Raman shift of 267 $cm^{-1}$ is shown (see the region indicated by the red square in Fig. 2a showing region of Raman investigations). Variations in the intensity of this mode are partially correlated with the thickness of the $MoTe_2$ layers on GaAs [31]. This is clearly seen in Fig. 3a, where the lowest intensity is recorded on the cobalt layer (spectrum 10) and the highest intensity on 1 L (regions R1, R2: spectra 05, 06), and in distorted region (spectrum 04). However, although the 267 $cm^{-1}$ Raman mode exhibits some sensitivity to the thickness, distorted and single layer $MoTe_2$ cannot be clearly distinguished, as the corresponding spectra (04, 05 and 06, respectively in Fig. 3a) show no significant intensity differences at this Raman shift. In contrast, differences between these areas become apparent in the map recorded at 124 $cm^{-1}$ shown in Fig. 3c, which we attribute to the degradation of $MoTe_2$ accompanied by the formation of metallic Te. We discuss this further below.

### *2.4.2. Raman spectra for different thicknesses of $MoTe_2$*

The Raman spectra recorded from the distorted $MoTe_2$ area are similar to those of GaAs (compare spectrum 04 with spectra 01–03 in Fig. 3a), but they also closely resemble spectra recorded from 1 L thick regions (see spectra 05 and 06 in Fig. 3a). This similarity, together with AFM and KPM data showing clear differences in height and CPD values (see SI Fig. S1), makes the interpretation of this region ambiguous. Particular caution is required in the low-wavenumber range, because some spectra were affected by weak instrumental artefacts related to residual laser-line reflections. These contributions were removed numerically, as described in Methods, by subtracting Gaussian components with a fixed intensity ratio. Importantly, after this correction, a pronounced feature at approximately 124 $cm^{-1}$ remains in the distorted region. We therefore associate this mode with the possible presence of metallic Te [30,31,44–46], although a contribution from the GaAs substrate or from strongly distorted $MoTe_2$ cannot be fully excluded. Consequently, we suggest that this region may correspond either to Te-rich nanostructures directly formed on GaAs or to a distorted $MoTe_2$ layer exhibiting signatures of Te segregation.

The character of the Raman spectra changes in the 4 L region (spectrum 07), where modes at 216 and 238 $cm^{-1}$ start to develop. For the 9 L thick film (spectra 08 and 09 recorded in R3 and R4 regions), these modes evolve into a broad multimode feature with several local maxima located in the range 200–240 $cm^{-1}$ (see spectra 07-09 in Fig. 3a). Another mode appearing in the 4 L region is located at 173 $cm^{-1}$ and for thicker films it shifts to 172 $cm^{-1}$, showing a considerable increase in intensity. A thickness-dependent shift of this mode (identified as the 2H-$MoTe_2$ $A_{1g}$ mode) has been reported previously, although it is typically observed to shift toward higher wavenumbers with increasing thickness [84–86]. Interestingly, multimode features at lower wavenumbers are occasionally observed and are attributed to Davydov components related to in-phase and out-of-phase interlayer interactions [86]. Therefore, the interpretation of both groups of modes – those in the 200–240 $cm^{-1}$ range and the mode near 171 $cm^{-1}$ remains nontrivial, particularly when strain and structural defects in the grown films are considered.

Weaker modes located at approx. 110–113 $cm^{-1}$, 100 $cm^{-1}$ and 136 $cm^{-1}$ can be observed in the 4 L and 9 L regions (spectra 07–09). Moreover, these modes are also present on the peeled off layers, which indicates that their origin is not related to GaAs, at least for thicker films (see spectrum 11 in Fig. 3a). One of the potential explanations might be formation of ultrasmall Te nanostructures [44] although a contribution from a dense grain boundary network, as suggested by the STM results, cannot be excluded.

#### *2.4.3. Raman spectra of $MoTe_2$ protected by a capping layer*

The majority of Raman investigations on 2D materials are conducted in air or after air exposure, and so it is in our case. Since this does not guarantee a clean surface, we evaluated the feasibility of performing Raman measurements through a metallic capping layer, which in our study is a 17 nm thick cobalt film (see spectrum 10 in Fig. 3a). Three distinct maxima can be seen in this spectrum, located at 201, 217, and 233 $cm^{-1}$. Another characteristic region is the energy range of 150-185 $cm^{-1}$, in which two distinct modes exist, located at 172 $cm^{-1}$ and 159 $cm^{-1}$ (the latter associated with GaAs). Overall, this spectrum closely resembles that recorded from the 9 L thick region, indicating that Raman measurements through the capping layer remain feasible.

#### *2.4.4. Simulation of Raman spectra of encapsulated $MoTe_2$*

The primary drawback of conducting measurements through the capping layer is a threefold signal attenuation observed for the 17 nm-thick cobalt layer. To understand qualitatively what thicknesses of the metallic capping layer can allow Raman measurements, we performed simulations using the transfer matrix method (TMM) code developed by S.J. Barnes [87] with modifications, which include the Raman scattering process [31] (see Fig. 3d for a cartoon showing the Raman scattering process). In short, the laser light illuminates the sample and multiple internal reflections in the metallic capping layer and $MoTe_2$ occur. Some of this light is absorbed in $MoTe_2$, leading to the Raman scattering, which after multiple internal reflections, is recorded outside the sample.

Results for nickel and cobalt, which we used in our experiments, are shown in Fig. 3e and Fig. 3f, respectively. The model used here suggests that cobalt is slightly better because of its lower extinction coefficient [55,88] and consequently, the overall obtained intensities of light are larger for Co than for Ni. The main conclusion from these simulations is that the thickness of the metallic layer should be lower than 20 nm, which guarantees sufficient Raman signal to conduct measurements. Moreover, materials characterized by a low extinction coefficient are better for capping layers because, in general, one might expect larger intensities of a measured signal.

#### *2.4.5. Raman spectroscopy on delaminated $MoTe_2$*

Finally, we investigate $MoTe_2$ after delamination, i.e., on the back side of the capping layer after it was peeled off. This experiment was conducted on samples with a nominal thickness of 4 L (covered by a Ni capping layer) [48] and was carried out in a protective atmosphere. The results recorded on the delaminated capping layer are shown in Fig. 3a – spectrum 11. In general, they are similar to the spectra recorded on $MoTe_2$/GaAs i.e., a broad multi-mode maximum is observed in the range 200-240 $cm^{-1}$ with the main maximum at 237 $cm^{-1}$, and few other at 103, 111, 137, and 173 $cm^{-1}$ (compare spectra 11 and 10 in Fig. 3c). Two other Raman modes can be seen, located at 266 and approx. 280 $cm^{-1}$, which for the samples grown on GaAs were obscured by its TO modes (located at 267 and 291 cm$^{-1}$). Note that the shape of the obtained spectrum is slightly different from the one recorded after the transfer of $MoTe_2$ grown on GaAs to SiO2 [73]. The reason could be the degradation of the layer in the past experiments [73], which in our case is minimal due to experiments conducted in the protective atmosphere [48].

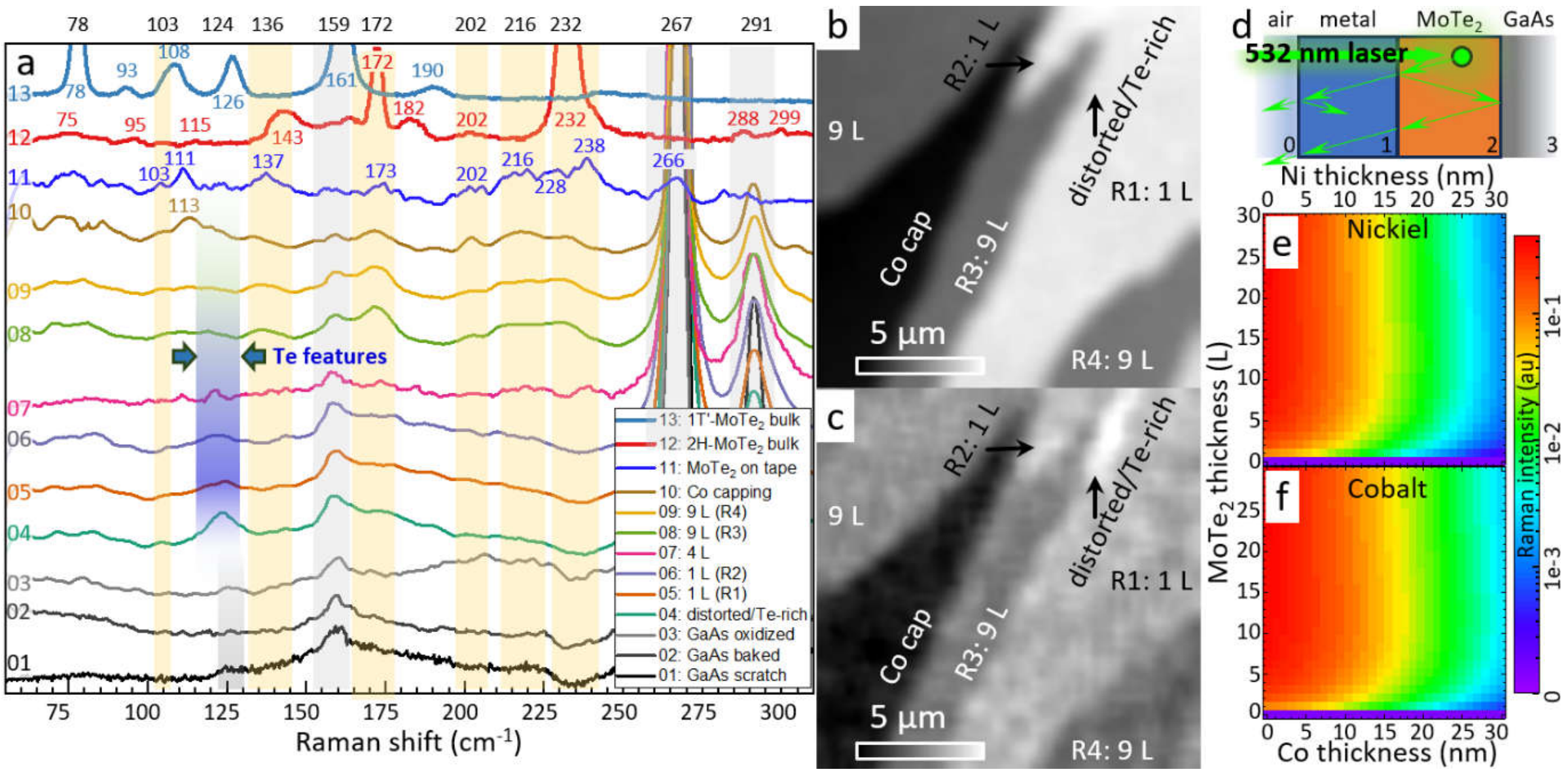

*Fig. 3. (a) Individual Raman spectra of GaAs (spectra 01-03), $MoTe_2$ extracted from different regions of the sample including distorted region, 1 L, 9 L thick regions and metallic capping layer (spectra 05-10), $MoTe_2$ exfoliated on adhesive tape (spectrum 11) and bulk 2H-$MoTe_2$ and 1T'-$MoTe_2$ (spectra 12 and 13, respectively). Grey and yellow marked ranges correspond to GaAS and grown $MoTe_2$ thick films respectively. Region indicated in blue corresponds to main Raman Te-related mode. Raman mapping at (b) 267 cm$^{-1}$ mode for GaAs and (c) 124 cm$^{-1}$. (d) Cartoon showing Fabry-Pérot resonance in thin layers. Simulation of Raman signal intensity as a function of (e) Ni and (f) Co capping layer thickness for different thicknesses of $MoTe_2$.*

## 2.5. Theoretical calculations

Finally, we present here a DFT theoretical analysis of the H and T' polytypes for thin films of $MoTe_2$. Since the processes described here are very complex (oxidation, grain boundaries, distorted twin domains, twists, etc.), we decided on a simplistic description that focuses on qualitative description. Namely, Janus structures [89] of H- and T' polytypes in which the top tellurium atoms are replaced by oxygen. To simulate surface oxidation in thicker layers, the oxygen atoms are substituted only in the top monolayer, as shown in *Fig. 4*a and *Fig. 4*b. It should be noted that to keep the similarity between the two systems in our numerical studies, we decided to calculate the 2H and 2T' polytypes as in other papers [90,91].

### *2.5.1. Band gaps and DOS*

The results of our calculations for the 2H-$MoTe_2$ polytype clearly indicate a bandgap width dependence on the number of layers as shown in *Fig. 4*c. The band gap width varies between 1.073 eV for 1 L through 0.611 eV for 5 L to 0.588 eV in the bulk (bulk value is estimated using a naive extrapolation by exponential decay, see also density of states – DOS shown in Fig. 4d). These values are different from the experimental ones (1.0 eV at 1 nm and 0.9 eV at 10 nm [23]) because of the typical underestimation of the gap in DFT [92]. In turn, the T' phase has metallic properties for a thickness of 1-5 L as evident from the calculated DOS (see Fig. 4e). Our calculations clearly show that Janus structures with oxygen substitution are characterized by metallic properties independently of the phase (see DOS in Fig. 4d and Fig. 4e, note zero DOS is indicated by dashed lines).

### 2.5.2. Work function

Our DFT calculations also provide the work function for H- and T' polytypes of $MoTe_2$ and MoTeO. As shown in Fig. 4f, these results show that the WF for T' does not depend on the number of layers and equals 4.48 eV. In contrast, WF for the H phase strongly depends on the thickness and varies from 4.78 eV for 1 L to 4.65 eV for 5 L. Assuming exponential decay with an infinite number of layers, the WF value reaches 4.64 eV for bulk.

The WF for $MoTe_2$ undergoes a rapid change after oxidation. The data shown in Fig. 4f (see also SI Fig. S3, where we compare two variants of monolayer oxide for polytype H and $MoO_2$ constructed by substitution of all Te atoms by oxygen) indicate that it varies depending on the side of the crystal, reaching much higher values on the oxygen side than on the tellurium-terminated one. This feature is universal, independent of the polytype. It clearly indicates that the oxygen-binding electrons have lower energies (i.e. stronger bonds) than those for tellurium, which results in the non-uniformity of WF for both sides of the crystal, i.e. a dipole is formed.

For oxidized H-$MoTe_2$, the WF decreases rapidly from 5.97 eV for 1 L to 5.56 eV for 5 L and to a similar value in the bulk. This trend points to a strong modification of the electronic structure and a weakening of the oxygen bonding between 1 L and 2 L. For thicker films, the WF becomes nearly constant, indicating that the oxygen-related bands have stabilized. A comparable evolution is observed for the tellurium-terminated side, where the WF increases from 3.35 eV for 1 L to 4.07 eV for 5 L and 4.10 eV in the bulk (see Fig. 4f).

Even greater changes in WF are observed for the T' polytype, both on the oxygen and tellurium sides. As can be seen in Fig. 4d on the oxygen (tellurium) side, WF changes from 5.54 eV (3.38 eV) for 1 L to 4.93 eV (4.44 eV) for 5 L and 4.84 eV (4.46 eV) for thick films. Such pronounced variations, compared to the metallic T'-$MoTe_2$, arise from a significant reduction in the O–Mo bond length, producing a configuration where parallel Mo–O rows are interconnected by Te bridges (see Fig. 4b). This configuration causes a non-equilibrium electron distribution and promotes the formation of dangling bonds, resulting in large changes of WF.

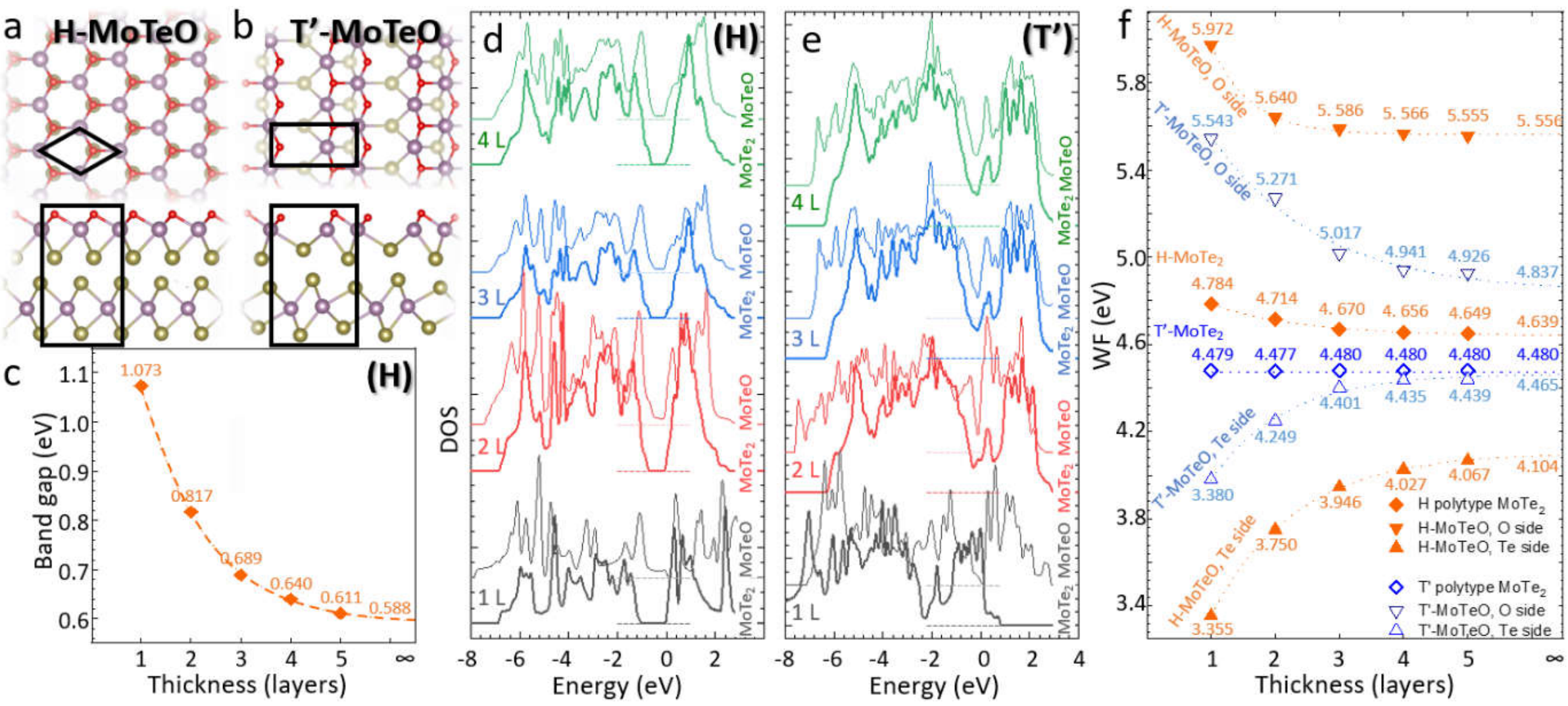

*Fig. 4. Ball and stick models showing (a) single layer of H-MoTeO and (b) T'-MoTeO on H- and T'-$MoTe_2$. (c) Band gap width calculated for 1-5 L of 2H-$MoTe_2$. Labels indicate calculated band gap widths. Evolution of density of states as a function of $MoTe_2$ thickness for (d) H- and (e) T' phases of $MoTe_2$ (thick lines) and MoTeO (thin lines). Horizontal dashed lines indicate zero of DOS in (d) and (e). (f) Work function for H- and T' phases of $MoTe_2$ and MoTeO. Labels show calculated WF.*

# 3. Discussion

## 3.1. Excess of Te in XPS and screening model

XPS analysis reveals a pronounced tellurium enrichment at the surface of $MoTe_2$ films grown on GaAs substrates. For samples exfoliated in UHV, the measured Te:Mo atomic ratio reaches approximately 4, which is above the stoichiometric value of 2.

Such stoichiometry is unrealistic for crystallographically stable $MoTe_2$; nevertheless, the result is robust, as it was consistently observed not only from the relative intensities of the Mo 3d and Te 3d states but also from other detectable core levels of these elements. Notably, a Te excess was also detected in our test measurements on bulk 2H-$MoTe_2$ crystals, further confirming the systematic nature of this observation. To better understand this discrepancy, we considered the possibility of systematic errors in the quantitative XPS analysis arising from the specific layered structure of $MoTe_2$. In this material, Te atoms occupy the outer planes of each layer, effectively surrounding the Mo atoms. This configuration may enhance the absorption of photoelectrons emitted from Mo relative to those from Te, thus biasing the stoichiometric ratios derived from XPS line intensities.

We introduce a simple qualitative screening model for XPS data from layered crystals. In this approach, the probability $P$ that a photoexcited electron escapes from a given depth $\alpha$ decreases exponentially:

$$P = e^{-a/\lambda},$$

where $\lambda$ is the inelastic mean free path of the electron in the material and the emission is at an angle normal to the surface. Assuming that individual atomic planes of Mo and Te atoms are parallel to the surface and distributed at subsequent depths, it is possible to describe the emission probabilities for particular states:

$$P_{Te} = \sum_{Te} e^{-a_{Te}/\lambda},\ P_{Mo} = \sum_{Mo} e^{-a_{Mo}/\lambda},$$

where $\alpha_{Te}$ and $\alpha_{Mo}$ enumerate the depths of successive Te and Mo layers and the summation is performed for a limited depth resulting from the nature of XPS studies. Since the measured XPS intensities are proportional to these probabilities, the ratio $P_{Te}$:$P_{Mo}$ provides a reasonable approximation of the experimentally determined Te:Mo ratio.

The exponential decay of probability described above follows the standard emission model, which assumes a homogeneous material - an assumption not fully valid for van der Waals structures. To better reflect the layered nature of $MoTe_2$, we assumed that effective electron scattering occurs mainly within the chemically bonded Mo-Te planes, while the van der Waals gaps contribute a little. In practice, this means that only the region around the Mo plane (≈0.3 nm within the 0.7 nm monolayer spacing) effectively scatters photoelectrons, further increasing the relative probability of detecting Te signals. Using a short mean free path of 0.6 nm, corresponding to an effective analysis depth of ~2.8 nm, our model yields a Te:Mo ratio of 2.9. Thus, XPS clearly overestimates the stoichiometry in layered structures. However, even under these extreme assumptions, the model cannot account for the nearly 4:1 ratio observed experimentally, indicating that our films indeed contain excess Te, though not to the extent suggested by a naive homogeneous model.

This discrepancy suggests that the epitaxial growth conditions, which typically involve a tellurium-rich flux, lead to an accumulation of excess Te near the surface, most likely in the form of Te clusters. During the deposition of the metallic capping layer, part of this excess Te may become trapped at the $MoTe_2$ - metal interface or between adjacent layers. After the capping is removed, a fraction of the Te-rich material remains on the $MoTe_2$ surface, giving rise to local deviations from stoichiometry.

STM measurements, however, do not directly reveal an excess of Te. Instead, imaging was deliberately performed away from visible surface contaminants, which likely excluded the regions containing Te clusters detected in XPS. In these cleaner areas, we observe a heterogeneous surface composed of atomic-resolution regions consistent with locally Te-deficient and stoichiometric $MoTe_2$. Together, these results indicate that the surface is chemically inhomogeneous: XPS captures the overall Te enrichment originating from clusters, while STM highlights local variations in the stoichiometry in cluster-free regions.

When such samples are exposed to air, tellurium reacts with oxygen to form Mo-Te-O species. In the early stages of oxidation, Te clusters that likely form at the surface and in near-surface regions are only partially oxidized, leading to a transient increase in the clean Te:Mo ratio detected by XPS. With prolonged exposure, however, these clusters undergo further oxidation, which enhances the O-Te signal and ultimately lowers the overall Te:Mo ratio.

### 3.2. Possible formation of Te-rich nanostructures

Previous studies have shown that Te nanostructures exhibit characteristic Raman features typically observed in the 120–128 $cm^{-1}$ range, accompanied by secondary modes near 140 $cm^{-1}$ and weaker signals around 103 $cm^{-1}$ [31,44–46]. The exact position of these modes is known to depend strongly on the morphology and size of Te nanostructures, and significant shifts have been reported for nanowires and other confined geometries in the range of 115-200 $cm^{-1}$ [44,47].

In our measurements, a mode located at 124 $cm^{-1}$, accompanied by weaker features at 143 and 103 $cm^{-1}$, is observed in distorted and 1 L $MoTe_2$ regions. However, its origin is not straightforward, as a mode near 124 $cm^{-1}$ is also detected on the GaAs substrate, with its intensity increasing upon oxidation (see spectra 01–03 in Fig. 3a). Two possible explanations can therefore be considered: (i) the formation of metallic Te or (ii) degradation of the substrate due to the limited thickness of the $MoTe_2$ layer. While neither scenario can be fully excluded, the results discussed below support the interpretation that Te-rich nanostructures may form locally.

Further support for this interpretation is provided by experiments showing that nanowire-like structures can form during Raman measurements themselves (see Fig. 5). These features appear predominantly in the monolayer regions, most clearly within the laser-scanned area (see Fig. 5a and the magnified view in Fig. 5b), but they can also extend slightly into the surrounding area (see inset in Fig. 5a). We attribute this effect to local heating induced by laser irradiation, which as we believe promotes the segregation of Te either released from degraded $MoTe_2$ or originating from excess Te present in the film in agreement with recent reports [30,31]. The segregated Te can then self-organize into elongated nanostructures. This process appears to be suppressed in thicker regions, likely due to more efficient heat dissipation and in consequence reduced mobility of Te atoms.

Additional support for the possible formation of metallic Te is provided by Raman measurements performed on 1 L regions exhibiting blister-like features that formed at the edges of the sample (see the optical image in the inset of SI Fig. S2b). Raman spectra recorded from these regions are generally characterized by a stronger mode at 124 $cm^{-1}$ compared to the surrounding 1 L areas. In some cases, this mode becomes even more intense and is accompanied by additional features at 143, 105, and 100 $cm^{-1}$, which is consistent with the formation of metallic tellurium [30,31,44–47].

For thicker films and in the peeled-off layers, weak modes at approximately 100 $cm^{-1}$, 110–113 $cm^{-1}$ and 136 $cm^{-1}$ are observed. Their presence in detached regions indicates that they are not related to the GaAs substrate. Instead, these features may be associated with metallic Te [44,47] and/or with structural defects, such as the dense grain boundary network observed in STM measurements.

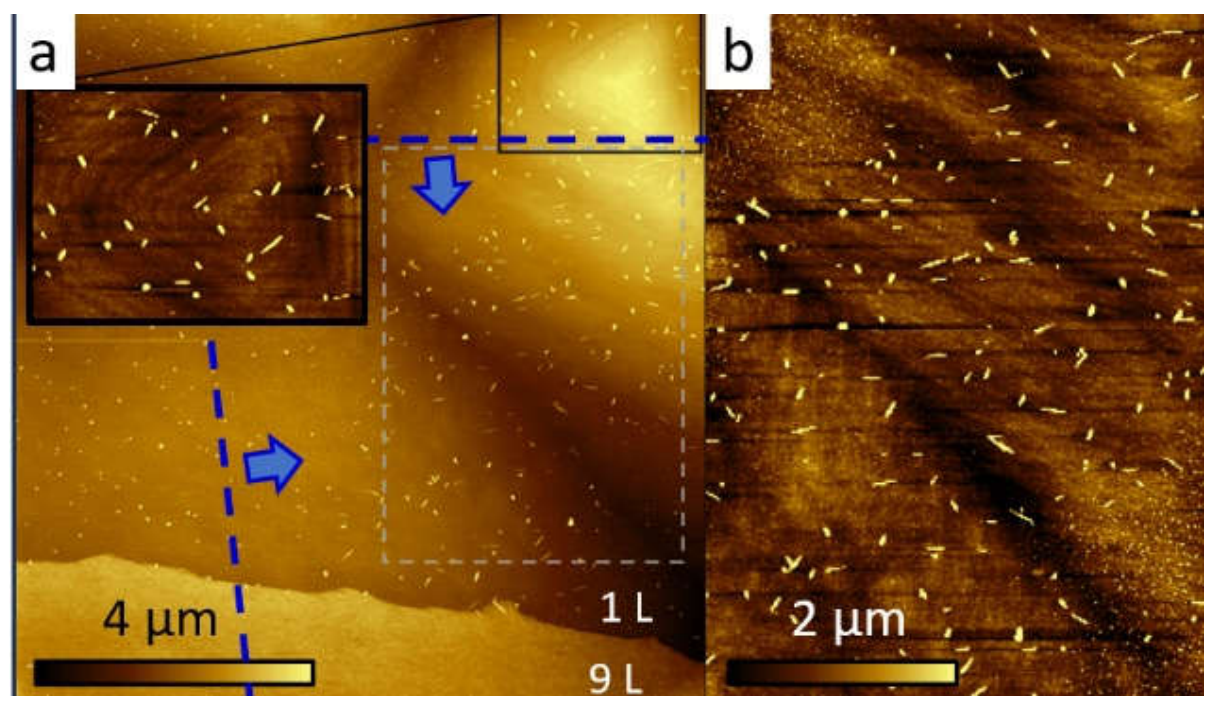

*Fig. 5. (a) AFM results recorded after Raman spectroscopy measurements of a 2.5 year old 2H-$MoTe_2$ sample. Thick blue dashed lines show edges, while arrows show the region in which Raman mapping was conducted. The inset in (a) shows nanowires formed outside the Raman-scanned region. (b) The magnified region indicated by a grey thin dashed line in (a) shows the formation of nanowires in the Raman scanned area.*

## 3.3. Non-stoichiometric signatures in Raman spectra of epitaxial $MoTe_2$

A detailed comparison of our Raman spectra with reference measurements performed on bulk 2H- and 1T′-$MoTe_2$ (see Fig. 3a, spectra 12 and 13, respectively) reveals a general resemblance to the 2H phase, yet several notable differences are observed. These include additional vibrational modes, as well as shifts in peak positions and changes in relative intensities. In our opinion, these spectral differences likely reflect structural and compositional deviations from the ideal stoichiometry. Previous studies have shown that lattice defects and non-stoichiometric compositions can significantly alter the Raman response of TMDCs [93–96]. The presence of excess tellurium and local inhomogeneities in surface orientation, confirmed by XPS and STM, suggests that the films investigated here represent a defective variant of 2H-$MoTe_2$.

Crucially, our spectra show no indication of features characteristic of the 1T′ phase, which excludes the possibility of the phase coexistence – a phenomenon often reported in $MoTe_2$ under strain, doping, or during growth [97–99]. All these observations support the interpretation that the $MoTe_2$ layers studied here are non-stoichiometric 2H-phase films, locally affected by Te-rich regions and defect-related structures.

## 3.4. Adhesion and interfacial bonding in- and between $MoTe_2$ films and Co, Ni capping layers

Our experiments reveal several important features related to the interlayer and interfacial bonding in epitaxial $MoTe_2$ films. First, the 1 L-thick base layer appears to be strongly bonded to the GaAs substrate. This is evidenced by the difficulty in identifying regions thinner than 1 L of $MoTe_2$, and by the fact that the cleavage plane during delamination typically occurs between the first and second layers.

We attribute this strong adhesion at the $MoTe_2$/GaAs interface to the high chemical reactivity of 2H-$MoTe_2$, especially in the presence of tellurium-related defects. These defect sites (either adatoms or vacancies) likely promote the formation of chemical bonds with the GaAs surface, anchoring the bottom layers and impeding delamination. In this sense, this resembles the covalent bonds of the buffer layer of graphene grown on SiC [100–102].

The absence of regions with intermediate thickness between 1 L and 9 L suggests the formation of a second preferred cleavage plane at the interface between the ninth and tenth layers. While the bottom layer is bonded to the substrate, the topmost one likely forms bonds with the metallic capping (see below for further discussion). As a result, the strength of interlayer interactions is not uniform across the film and is weakest near the interfaces, i.e., between the first and second, and the ninth and tenth layers. This suggests that interlayer coupling in $MoTe_2$ is governed by the balance between substrate adhesion and bonding to the metallic overlayer.

A time-dependent evolution of bonding is also observed in capped films. Immediately after the growth and encapsulation, the metallic capping layer can easily be removed using an adhesive tape. However, after just 24 hours, delamination becomes more difficult, yet still possible. After prolonged storage, the capping layer can no longer be mechanically removed. This behavior suggests the gradual formation of interfacial chemical bonds over time – both between $MoTe_2$ layers and between $MoTe_2$ and the metallic overlayer. The latter is plausible given that both nickel and cobalt are known to form chemical bonds with tellurium [103,104], and our films exhibit tellurium excess at the interface. As a result, the formation of mixed-metal bonds such as Co–Te–Mo or Ni–Te–Mo is possible, which hinders the removal of the protective layer. At the same time, additional bonds may form between adjacent $MoTe_2$ layers, initiated by the excess tellurium, overall non-stoichiometry of the film, and the presence of locally twisted domains observed by STM. Together, these factors can promote interlayer bonding, which prevents the formation of new cleavage planes and ultimately hinders the removal of the capping layer.

These findings shed light on the chemical interactions, stability, and protective role of Ni and Co capping layers on $MoTe_2$. While effective for short-term protection, their reactivity with tellurium limits their use in applications that require delayed delamination. Nevertheless, their lower vapor pressure and lower toxicity compared to elemental Te [105–107], which could be used as an alternative cap, makes them suitable for transient protection during sample transfer.

### 3.5. Oxidation and ageing

Our experiments reveal a strong dependence of CPD and WF on the film thickness. Theoretical DFT calculations predict a monotonic trend for pristine and uniformly oxidized films, with thinner layers exhibiting higher WF (i.e., lower CPD). In this scenario, one would expect $CPD_{nL} < CPD_{(n+1)L}$ for n representing thickness. However, our measurements show the opposite trend: $CPD_{9L} < CPD_{1L}$. This discrepancy suggests that oxidation in real samples does not follow a uniform or idealized model.

To investigate this further, we performed DFT calculations for several oxidized $MoTe_2$ related compounds, including monolayers of $MoTe_2O$, MoTeO, and $MoO_2$ (see SI Fig. S3). Each of these oxides exhibits a distinct work function, indicating that the nature of the oxide plays a critical role in the observed CPD values. Based on these results, we speculate that the level of oxidation is larger for thicker films and in consequence, the CPD is lower (compare $MoO_2$ and MoTeO in SI Fig. S3). We therefore propose that the experimental CPD behavior results from a combination of several factors: (i) intrinsic thickness dependence of the electronic structure, (ii) formation of different oxide species on films of different thicknesses, (iii) non-uniform oxidation depth in particular if 1 L thick films are compared to 9 L, (iv) potential degradation of the GaAs substrate under ultrathin $MoTe_2$ regions, which could affect local CPD through interfacial charge redistribution, and (v) excess of Te which forms oxides affecting simplistic theoretical picture. Supporting this interpretation, our sample-ageing experiments show a progressive increase in CPD relative to the cobalt capping layer across all samples, but with a markedly greater shift for 1 L films compared to the 9 L ones. This indicates that ultrathin films oxidize faster or via different mechanisms than thicker ones.

## 4. Conclusions

In this study, we aimed to shed light on the interaction between metallic capping layers and epitaxial 2H-$MoTe_2$ films, as well as to evaluate their feasibility for surface protection and optical transparency in Raman studies. Our findings demonstrate that thin Co and Ni overlayers effectively protect $MoTe_2$ against oxidation in the short term. However, due to the formation of interfacial chemical bonds – likely facilitated by tellurium adatoms or vacancies – their protective performance deteriorates over time. After several days, delamination becomes increasingly difficult, and recovery of a clean $MoTe_2$ surface is no longer possible.

We also find that the 1 L $MoTe_2$ layers exhibit strong adhesion to the GaAs substrate, likely due to enhanced interfacial reactivity and defect-related bonding. The abundance of exposed GaAs regions in AFM and the specific cleavage plane at 1 L further support this interpretation. These observations emphasize the importance of timing when delaminating encapsulated samples and have direct implications for the practical handling of air-sensitive 2D materials.

An additional observation is the local appearance of elongated nanostructures after laser exposure, which, together with Te-rich XPS signatures and selected Raman spectra, suggests laser-assisted Te segregation.

Our findings clearly show that $MoTe_2$ surfaces degrade rapidly upon air exposure. We observed a strong correlation between contact potential difference, film thickness, and the extent of oxidation. These experimental results are qualitatively supported by DFT calculations, which indicate that key electronic properties such as density of states and work function are strongly thickness-dependent for both pristine and oxidized forms of 1T′ and 2H phases. The observed discrepancies between the theory and the experiment suggest that different oxide species may form on layers of different thicknesses, influencing degradation pathways and surface energetics.

Finally, we demonstrate that thin metallic capping layers (<20 nm) serve a dual purpose: they protect the sensitive $MoTe_2$ surface from rapid degradation and simultaneously allow for semitransparent optical access. This capability enables Raman spectroscopy and potentially other optical techniques to be carried out in ambient conditions without compromising sample integrity – offering a practical solution for characterizing van der Waals materials under realistic constraints.

## Methods

### Growth

The growth of 2H-$MoTe_2$ layers was carried out using a dual-chamber molecular beam epitaxy system, comprising separate chambers dedicated to II–VI and III–V compound growth. High-purity molybdenum (99.995%) was evaporated from an electron beam source, while tellurium was supplied via a dual-filament Knudsen cell. Semi-insulating, epi-ready GaAs(111)B wafers served as the substrate material. Prior to $MoTe_2$ deposition, native oxides on the GaAs(111)B surface were thermally desorbed by heating the substrate to 600 °C within the II–VI chamber. $MoTe_2$ layers were then grown at a substrate temperature of 370 °C, with the tellurium source maintained at 385 °C, providing a highly Te-rich growth environment.

## Samples

Several $MoTe_2$/GaAs and reference samples were used in this study, differing in nominal $MoTe_2$ thickness, capping material, delamination procedure, ageing time, and measurement environment. To clarify the relation between individual datasets, figures, and sample histories, a complete overview is provided in SI Table S1. The main dataset was obtained from nominally 10 L $MoTe_2$ films capped with approximately 20 nm of Co, which were studied after decapping either in UHV or after exposure to air for times ranging from several hours to approximately 2.5 years. Additional measurements were performed on nominally 4 L $MoTe_2$ films protected with Ni capping layers and decapped either in UHV or in an argon atmosphere. Reference Raman spectra were collected from GaAs substrates not exposed to Te, as well as from freshly exfoliated bulk 2H-$MoTe_2$ and 1T′-$MoTe_2$ crystals. The term "fresh" refers to samples measured immediately after decapping or exfoliation in the indicated environment.

## XPS measurements

XPS measurements were performed at room temperature using an Omicron UHV system, with a base pressure of approximately $2 \times 10^{-10}$ mbar during data acquisition using a DAR 400 X-ray source operating with non-monochromatic Mg Kα radiation (1253.64 eV, 0.7 eV linewidth). A hemispherical energy analyzer (Phoibos 150, SPECS) equipped with a 2D CCD detector was employed, with the pass energy set to 20 eV. The overall energy resolution was approximately 1.0 eV. All the measurements were performed in normal emission geometry on thin film samples. Spectral analysis was carried out using CasaXPS software. All binding energies were calibrated with respect to the C 1s peak at 284.5 eV. Quantitative evaluation included Shirley background subtraction and peak fitting. Peak fitting was performed using mixed Gaussian–Lorentzian line shapes GL(50) for oxide peaks and Lorentzian asymmetric with tail dumping LF(α, β, w, m) for metallic peaks. The full width at half maximum (FWHM) and spectrum area parameters were constrained for the corresponding multiplet states. Relative atomic concentrations were estimated using relative sensitivity factors provided within CasaXPS. The binding energies, FWHM values and line shapes used for fitting the Mo 3d and Te 3d regions are summarized in SI Table S2. Based on repeated fits with varied background limits, allowed FWHM ranges and exclusion of analyzer transmission the uncertainty of the reported relative ratios is estimated to be within approximately 10–15%. Satellite features were included in the fitting where relevant.

## STM and LEED characterization

STM and LEED measurements were conducted in the UHV Omicron system at room temperature with a base pressure of $2 \times 10^{-10}$ mbar. STM imaging was performed using mechanically cut tips made from 90% Pt–10% Ir alloy wires (Goodfellow). For data acquisition, we used the dedicated MATRIX V3.3.2 software. Typical imaging conditions involved a bias voltage in the range of ±1 V and a tunneling current of 0.1-1.0 nA. Scan sizes and image-specific tunneling parameters are given in the corresponding figure captions. The resulting topographic data was processed and visualized with Gwyddion 2.59 [108].

The BDL800IR (OCI Vacuum Microengineering Inc.) device paired with a CCD camera was used to perform LEED measurements, while WSxM 5.0 [109] was used to process the results. The data were acquired in a range of 50-100 eV. Diffraction patterns were recorded at near-normal incidence and were used to assess surface crystallinity and long-range order. The exact beam spot size was not calibrated, but it was smaller than the analyzed homogeneous surface region.

## AFM and KPM measurements

AFM and KPM measurements were conducted using the NTEGRA PNL control system from NT-MDT under ambient conditions. For the probes, silicon TipsNano NSG01 were used. Tapping mode was used to acquire topographical images with a scanning frequency ranging between 0.1 and 1.0 Hz, depending on the scan size and resolution. KPFM measurements were carried out in S/Cont. II pass AM-SKM mode. Typical images were recorded with a resolution of 256 × 256 or 512 × 512 pixels over scan areas of 1-6400 $\mu m^2$. Because different scan sizes were used, the amplitude setpoint was adjusted individually; typical values were 60–80% of the free oscillation amplitude. Measurements were performed at room temperature in ambient air. Relative humidity was not actively controlled. Layer heights were extracted from line profiles after plane subtraction; the uncertainty was estimated from several profiles across equivalent step edges. Data analysis was conducted using Gwyddion 2.59 software [108], including standard image flattening for the topographical images.

## Raman spectroscopy

All Raman experiments were conducted in a protective argon atmosphere at room temperature using a Raman microscopy system described elsewhere [48,110,111] in a low-vibration laboratory environment [112]. In short, we used the imaging Cherny-Turner monochromator-spectrograph MS5204i (SOL Instruments) with a focal length of 520 mm, combined with a CCD camera containing HS 101H-2048/122-HR2 (Hamamatsu) sensor cooled down to -40 °C. The monochromator-spectrograph is equipped with a 4-position motorized grating turret with gratings 1800 l/mm (used in all experiments), 600 l/mm, 150 l/mm, and 75 l/mm Echelle optimized for a 532 nm laser (Compass C215M-50 from Coherent Inc.) operated at a power of ~4.6 mW. The measurements were performed using a 50×, NA = 0.55 objective LD EC Epiplan-Neofluar 50×/0.55 HD DIC M27 from Zeiss. The acquisition window of approximately 820 $cm^{-1}$, typically covering -91–730 $cm^{-1}$ range was used with a spectral resolution of approximately 0.4 $cm^{-1}$. The wavenumber scale was calibrated using the position of the reflected laser light and GaAs mode at 267 $cm^{-1}$ when available. All the acquired data were processed by subtracting the time-averaged detector white-noise background to improve the signal-to-noise ratio.

Raman maps and spectra 04-06, 08-10 shown in this work were recorded on a sample aged approximately 2.5 years as a grid measurement recorded over a 20×20 $\mu m^2$ region using a 50×50 mesh of points, corresponding to a step size of 400 nm. The spectra were acquired once per point, with an acquisition time of 12 s per point. These data were analyzed using Fiji ImageJ software [113]. A moving median (window size of 11 points) and Gaussian blur (radius set to 1.5 points) were applied to all spectra and spatial maps to reduce noise and remove spike-like artefacts. All spectra were extracted as area-averaged signals from selected regions containing 24, 472, 25, 97, 76, and 268 pixels for spectra 04-06, 08-10, respectively.

Spectra 07 (4 L $MoTe_2$ covered by a nominally 40 nm thick Ni capping layer) and 11 ($MoTe_2$ on tape; nominally 120 nm thick Ni capping layer grown on 4 L $MoTe_2$ was exfoliated in Ar and the remaining $MoTe_2$ on the cap was measured) were acquired as point measurements on samples exfoliated and measured in an argon atmosphere. Spectrum 07 was recorded for 15 s and averaged over 2910 acquisitions and spectrum 11 was recorded for 300 s and averaged over 100 individual measurements.

Reference spectra 01 (GaAs scratched in an argon atmosphere) were recorded using an acquisition time of 5 s and 900 averages (taken from a 30×30 mesh of points). Spectrum 02 (backside of GaAs after annealing in UHV at ~700 °C for 20 min) was recorded using an acquisition time of 15 s and using a 40×40 grid, while 03 (heavily oxidized GaAs) for 60 s and using 100 averages. Spectra 12 (2H-$MoTe_2$) and 13 (1T'-$MoTe_2$) were recorded immediately after exfoliation of bulk crystals in argon (acquisition time 31 s and 60 s, 1600 and 30 averages, respectively).

Spectrum 14 (1 L blister-like feature) was obtained from an approx. 5-year-old sample in a region formed near the edge of the sample after breaking the crystal. The spectrum was extracted from a Raman grid measurement recorded over a 20×20 $\mu m^2$ region using a 100×100 mesh (step size 200 nm, acquisition time 3 s and no averages per point) by averaging approx. 30 individual measurements.

The Raman spectra (02, 04–11) were affected by background contributions, including weak photoluminescence and reflections of the laser line located at approximately 80 and 120 $cm^{-1}$. These artefacts were removed numerically. A constant intensity ratio of 1:0.24 between the two artefacts was assumed for most spectra. For spectra recorded through the capping layer and on adhesive tape, the relative intensity of the 120 $cm^{-1}$ feature increased up to ~50%, and this was accounted for during the correction procedure.

### DFT calculations

First-principles calculations for all the structures considered were performed using the QUANTUM ESPRESSO suite [114,115], implementing the DFT formalism using a plane wave basis [116]. Model systems were constructed using a slab geometry, with a vacuum of at least 19 Å used to separate the periodic images. To compensate for the surface dipole in the slab geometry for $MoTe_2$-MoTeO structures, a method introduced by Ref. [117] was applied. The exchange-correlation potential of the revised Perdew–Burke–Ernzerhof (PBEsol) type [118] and projected augmented wave (PAW) approach [119] were utilized. Scalar relativistic pseudopotentials (for lattice constant relaxation) and full relativistic pseudopotentials (for the later stages of calculations) with non-linear core correction taken from PSLibrary [120] were used. The energy cut-off for wavefunctions and charge density calculations amounted to 53 and 402 Ry, respectively. To account for the long-range van der Waals interactions, a semi-empirical approach following S. Grimme et al. [121] was applied. For the Brillouin zone sampling, a Methfessel-Paxton smearing of 0.001 Ry was used [122]. Prior to self-consistent and non-self-consistent calculations of the total energy of the structure, a relaxation of the lattice constants and the atomic positions was performed. For the DOS prediction, non-self-consistent calculations were performed with integration over the Brillouin zone according to Ref. [123].

**CRediT authorship contribution statement**

**Wojciech Ryś**: Investigation, Writing - Review & Editing.

**Iaroslav Lutsyk**: Investigation, Resources, Validation.

**Michał Piskorski**: Software, Methodology, Investigation, Formal analysis.

**Maxime Le Ster**: Writing - Review & Editing, Validation.

**Maciej Rogala**: Writing - Review & Editing, Validation.

**Paweł Dąbrowski**: Writing - Review & Editing, Validation.

**Paweł Krukowski**: Writing - Review & Editing, Validation.

**Katarzyna Ranoszek-Soliwoda**: Writing - Review & Editing, Validation

**Jarosław Grobelny**: Investigation, Writing - Review & Editing, Validation

**Zuzanna Ogorzałek-Sory**: Methodology, Resources, Validation.

**Bartłomiej Seredyński**: Methodology, Resources, Validation.

**Wojciech Pacuski**: Methodology, Resources, Validation.

**Janusz Sadowski**: Methodology, Resources, Validation.

**Marta Gryglas-Borysiewicz**: Methodology, Resources, Validation.

**Karol Szałowski**: Software, Methodology, Resources, Writing - Review & Editing.

**Paweł J. Kowalczyk**: Writing – original draft, Visualization, Software, Methodology, Data Curation, Formal analysis, Funding acquisition, Conceptualization.

**Acknowledgements**

This work was financially supported by the National Science Centre (Poland) under grant 2018/31/B/ST3/02450. As part of the 2021/41/N/ST5/04206 grant, a 2H-$MoTe_2$ layers were grown on a GaAs (111)B substrate using the MBE technique. This research was also supported by the Polish Ministry of Science and Higher Education under contract no. 2025/WK/01. Portions of the manuscript text were corrected using AI-based tools to improve grammar, clarity and readability of the manuscript.

**Data availability.** The data that support the findings of this study are available from the corresponding author upon reasonable request.

## References

1. Duan, X.; Zhang, H. Introduction: Two-Dimensional Layered Transition Metal Dichalcogenides. *Chem. Rev.* **2024**, *124*, 10619–10622, doi:10.1021/acs.chemrev.4c00586.
2. Li, J.; Yang, R.; Li, R.; Grigoropoulos, C.P. Exciton Dynamics in 2D Transition Metal Dichalcogenides. *Adv. Opt. Mater.* **2025**, *13*, doi:10.1002/adom.202403137.
3. Gupta, S.; Zhang, J.-J.; Lei, J.; Yu, H.; Liu, M.; Zou, X.; Yakobson, B.I. Two-Dimensional Transition Metal Dichalcogenides: A Theory and Simulation Perspective. *Chem. Rev.* **2025**, *125*, 786–834, doi:10.1021/acs.chemrev.4c00628.
4. Hussain, S.; Zhao, S.; Zhang, Q.; Tao, L. Comparative Analysis of Thin and Thick $MoTe_2$ Photodetectors: Implications for Next-Generation Optoelectronics. *Nanomaterials* **2024**, *14*, 1804, doi:10.3390/nano14221804.

5. Zhao, W.; Zhou, X.; Yan, D.; Huang, Y.; Li, C.; Gao, Q.; Moras, P.; Sheverdyaeva, P.M.; Rong, H.; Cai, Y.; et al. Synthesis and Electronic Structure of Atomically Thin 2H-MoTe2. *Nanoscale* **2025**, *17*, 10901–10909, doi:10.1039/D4NR05191B.

6. Fang, M.; Gu, H.; Guo, Z.; Liu, J.; Huang, L.; Liu, S. Temperature and Thickness Dependent Dielectric Functions of $MoTe_2$ Thin Films Investigated by Spectroscopic Ellipsometry. *Appl. Surf. Sci.* **2022**, *605*, 154813, doi:10.1016/j.apsusc.2022.154813.

7. Kumar, N.; Bhatt, K.; Kandar, S.; Rana, G.; Bera, C.; Kapoor, A.K.; Singh, R. Optical Properties of Nanoscale-Thick 2H and 1T′ $MoTe_2$ Films via Spectroscopic Ellipsometry: Implications for Optoelectronic Devices. *ACS Appl. Nano Mater.* **2024**, *7*, 23834–23841, doi:10.1021/acsanm.4c04325.

8. Liu, X.; Islam, A.; Guo, J.; Feng, P.X.-L. Controlling Polarity of MoTe 2 Transistors for Monolithic Complementary Logic *via* Schottky Contact Engineering. *ACS Nano* **2020**, *14*, 1457–1467, doi:10.1021/acsnano.9b05502.

9. Qi, D.; Han, C.; Rong, X.; Zhang, X.-W.; Chhowalla, M.; Wee, A.T.S.; Zhang, W. Continuously Tuning Electronic Properties of Few-Layer Molybdenum Ditelluride with in Situ Aluminum Modification toward Ultrahigh Gain Complementary Inverters. *ACS Nano* **2019**, *13*, 9464–9472, doi:10.1021/acsnano.9b04416.

10. Hidding, J.; Cordero-Silis, C.A.; Vaquero, D.; Rompotis, K.P.; Quereda, J.; Guimarães, M.H.D. Locally Phase-Engineered $MoTe_2$ for Near-Infrared Photodetectors. *ACS Photonics* **2024**, doi:10.1021/acsphotonics.4c00896.

11. Wang, Y.; Zhang, M.; Xue, Z.; Chen, X.; Mei, Y.; Chu, P.K.; Tian, Z.; Wu, X.; Di, Z. Atomistic Observation of the Local Phase Transition in MoTe 2 for Application in Homojunction Photodetectors. *Small* **2022**, *18*, doi:10.1002/smll.202200913.

12. Ghimire, M.K.; Ji, H.; Gul, H.Z.; Yi, H.; Jiang, J.; Lim, S.C. Defect-Affected Photocurrent in MoTe 2 FETs. *ACS Appl. Mater. Interfaces* **2019**, *11*, 10068–10073, doi:10.1021/acsami.9b00050.

13. Shi, J.; Bie, Y.-Q.; Zong, A.; Fang, S.; Chen, W.; Han, J.; Cao, Z.; Zhang, Y.; Taniguchi, T.; Watanabe, K.; et al. Intrinsic 1T′ Phase Induced in Atomically Thin 2H-MoTe2 by a Single Terahertz Pulse. *Nat. Commun.* **2023**, *14*, 5905, doi:10.1038/s41467-023-41291-w.

14. Ueno, K.; Fukushima, K. Changes in Structure and Chemical Composition of $\alpha$-$MoTe_2$ and $\beta$-$MoTe_2$ during Heating in Vacuum Conditions. *Applied Physics Express* **2015**, *8*, 095201, doi:10.7567/APEX.8.095201.

15. Tsipas, P.; Fragkos, S.; Tsoutsou, D.; Alvarez, C.; Sant, R.; Renaud, G.; Okuno, H.; Dimoulas, A. Direct Observation at Room Temperature of the Orthorhombic Weyl Semimetal Phase in Thin Epitaxial $MoTe_2$. *Adv. Funct. Mater.* **2018**, *28*, 1802084, doi:10.1002/adfm.201802084.

16. Ahmed, F.; Rodríguez-Fernández, C.; Fernandez, H.A.; Zhang, Y.; Shafi, A.M.; Uddin, M.G.; Cui, X.; Yoon, H.H.; Mehmood, N.; Liapis, A.C.; et al. Deterministic Polymorphic Engineering of $MoTe_2$ for Photonic and Optoelectronic Applications. *Adv. Funct. Mater.* **2023**, *33*, 2302051, doi:10.1002/adfm.202302051.

17. Ryu, H.; Lee, Y.; Jeong, J.H.; Lee, Y.; Cheon, Y.; Watanabe, K.; Taniguchi, T.; Kim, K.; Cheong, H.; Lee, C.; et al. Laser-Induced Phase Transition and Patterning of HBN-Encapsulated $MoTe_2$. *Small* **2023**, *19*, 2205224, doi:10.1002/smll.202205224.

18. Zhang, Q.; Zhang, Y.; Gao, G.; Zhang, S. Potential-Driven Semiconductor-to-Metal Transition in Monolayer Transition Metal Dichalcogenides. *Adv. Funct. Mater.* **2023**, *33*, doi:10.1002/adfm.202208736.

19. Deng, Y.; Zhao, X.; Zhu, C.; Li, P.; Duan, R.; Liu, G.; Liu, Z. $MoTe_2$: Semiconductor or Semimetal? *ACS Nano* **2021**, *15*, 12465–12474, doi:10.1021/acsnano.1c01816.

20. Duerloo, K.-A.N.; Li, Y.; Reed, E.J. Structural Phase Transitions in Two-Dimensional Mo- and W-Dichalcogenide Monolayers. *Nat. Commun.* **2014**, *5*, 4214, doi:10.1038/ncomms5214.

21. Gong, C.; Zhang, Y.; Chen, W.; Chu, J.; Lei, T.; Pu, J.; Dai, L.; Wu, C.; Cheng, Y.; Zhai, T.; et al. Electronic and Optoelectronic Applications Based on 2D Novel Anisotropic Transition Metal Dichalcogenides. *Advanced Science* **2017**, *4*, 1700231, doi:10.1002/advs.201700231.

22. Huang, J.; Deng, K.; Liu, P.; Wu, C.; Chou, C.; Chang, W.; Lee, Y.; Hou, T. Large-Area 2D Layered $MoTe_2$ by Physical Vapor Deposition and Solid-Phase Crystallization in a Tellurium-Free Atmosphere. *Adv. Mater. Interfaces* **2017**, *4*, doi:10.1002/admi.201700157.

23. Jung, E.; Park, J.C.; Seo, Y.-S.; Kim, J.-H.; Hwang, J.; Lee, Y.H. Unusually Large Exciton Binding Energy in Multilayered 2H-MoTe2. *Sci. Rep.* **2022**, *12*, 4543, doi:10.1038/s41598-022-08692-1.

24. Cheng, F.-J.; Lou, C.-C.; Chen, A.-X.; Wei, L.-X.; Liu, Y.; Deng, B.-Y.; Li, F.; Wang, Z.; Xue, Q.-K.; Ma, X.-C.; et al. Imaging Sublattice Cooper-Pair Density Waves in Monolayer 1T-$MoTe_2$. *Phys. Rev. Lett.* **2025**, *135*, 166201, doi:10.1103/x6k9-wgk9.

25. Zakhidov, D.; Rehn, D.A.; Reed, E.J.; Salleo, A. Reversible Electrochemical Phase Change in Monolayer to Bulk-like $MoTe_2$ by Ionic Liquid Gating. *ACS Nano* **2020**, *14*, 2894–2903, doi:10.1021/acsnano.9b07095.

26. Kim, C.; Issarapanacheewin, S.; Moon, I.; Lee, K.Y.; Ra, C.; Lee, S.; Yang, Z.; Yoo, W.J. High-Electric-Field-Induced Phase Transition and Electrical Breakdown of $MoTe_2$. *Adv. Electron. Mater.* **2020**, *6*, doi:10.1002/aelm.201900964.

27. Khan, R.; Rehman, N.U.; Kalluri, S.; Elumalai, S.; Saritha, A.; Fakhar-e-alam, M.; Ikram, M.; Abdullaev, S.; Rahman, N.; Sangaraju, S. 2D $MoTe_2$ Memristors for Energy-Efficient Artificial Synapses and Neuromorphic Applications. *Nanoscale* **2025**, *17*, 13174–13206, doi:10.1039/D5NR01509J.

28. Shinde, P. V.; Hussain, M.; Moretti, E.; Vomiero, A. Advances in Two-dimensional Molybdenum Ditelluride ($MoTe_2$): A Comprehensive Review of Properties, Preparation Methods, and Applications. *SusMat* **2024**, *4*, doi:10.1002/sus2.236.

29. Kuo, Y.-Z.; Liu, M.-J.; Sino, P.A.; Lai, P.-C.; Chung, C.-C.; Hsu, Y.-C.; Yang, T.-Y.; Cyu, R.-H.; Chuang, F.-C.; Kuo, H.-C.; et al. Direct Formation of Stable 1T′ Molybdenum Telluride ($MoTe_2$ ) by Laser Annealing Processes as Robust Contacts for High-Performance Molybdenum Disulfide ($MoS_2$ ) Field Effect Transistors. *ACS Appl. Mater. Interfaces* **2025**, doi:10.1021/acsami.5c12868.

30. Varjamo, S.-T.; Edwards, C.; Zhou, Y.; Fang, R.; Hosseini Shokouh, S.H.; Sun, Z. Optical Modification of TMD Heterostructures. *Nano Lett.* **2025**, *25*, 4379–4385, doi:10.1021/acs.nanolett.4c06512.

31. Piskorski, M.; Ryś, W.; Dunal, R.; Le Ster, M.; Krempiński, P.; Radecki, Ł.; Nadolska, A.; Toczek, K.; Przybysz, P.; Lutsyk, I.; et al. Survey of Raman Spectroscopy and Fabry–Pérot Interference in Thin Flakes of 2H-$MoS_2$, 1T-$TaS_2$, 1T'-$MoTe_2$, $T_d$-$WTe_2$, and α-$MoO_3$. *The Journal of Physical Chemistry C* **2025**, *129*, 18092–18111, doi:10.1021/acs.jpcc.5c04807.

32. Tan, Y.; Luo, F.; Zhu, M.; Xu, X.; Ye, Y.; Li, B.; Wang, G.; Luo, W.; Zheng, X.; Wu, N.; et al. Controllable 2H-to-1T′ Phase Transition in Few-Layer $MoTe_2$. *Nanoscale* **2018**, *10*, 19964–19971, doi:10.1039/C8NR06115G.

33. Tan, Y.; Luo, F.; Zhu, M.; Xu, X.; Ye, Y.; Li, B.; Wang, G.; Luo, W.; Zheng, X.; Wu, N.; et al. Correction and Removal of Expression of Concern: Controllable 2H-to-1T′ Phase Transition in Few-Layer $MoTe_2$. *Nanoscale* **2019**, *11*, 23498–23501, doi:10.1039/C9NR90258A.

34. Kim, T.; Park, H.; Joung, D.; Kim, D.; Lee, R.; Shin, C.H.; Diware, M.; Chegal, W.; Jeong, S.H.; Shin, J.C.; et al. Wafer-Scale Epitaxial 1T′, 1T′–2H Mixed, and 2H Phases $MoTe_2$ Thin Films Grown by Metal–Organic Chemical Vapor Deposition. *Adv. Mater. Interfaces* **2018**, *5*, 1800439, doi:10.1002/admi.201800439.

35. Wang, J.; Zhang, M.; Zhou, Z.; Li, L.; Song, Q.; Yan, P. Room-Temperature $MoTe_2$/InSb Heterostructure Large-Area Terahertz Detector. *Infrared Phys. Technol.* **2024**, *137*, 105190, doi:10.1016/j.infrared.2024.105190.

36. Cho, S.; Kim, S.; Kim, J.H.; Zhao, J.; Seok, J.; Keum, D.H.; Baik, J.; Choe, D.-H.; Chang, K.J.; Suenaga, K.; et al. Phase Patterning for Ohmic Homojunction Contact in $MoTe_2$. *Science (1979).* **2015**, *349*, 625–628, doi:10.1126/science.aab3175.

37. Yuan, S.; Luo, X.; Chan, H.L.; Xiao, C.; Dai, Y.; Xie, M.; Hao, J. Room-Temperature Ferroelectricity in $MoTe_2$ down to the Atomic Monolayer Limit. *Nat. Commun.* **2019**, *10*, 1775, doi:10.1038/s41467-019-09669-x.

38. Hussain, S.; Patil, S.A.; Vikraman, D.; Mengal, N.; Liu, H.; Song, W.; An, K.-S.; Jeong, S.H.; Kim, H.-S.; Jung, J. Large Area Growth of $MoTe_2$ Films as High Performance Counter Electrodes for Dye-Sensitized Solar Cells. *Sci. Rep.* **2018**, *8*, 29, doi:10.1038/s41598-017-18067-6.

39. Fernández García, A.; Torres-Costa, V.; de Melo, O.; Agulló-Rueda, F.; Castro, G.R.; Manso Silvan, M. Growth of Out-of-Plane Standing $MoTe_{2(1-x)}Se_{2x}$/$MoSe_2$ Composite Flake Films by Sol–Gel Nucleation of MoOy and Isothermal Closed Space Telluro-Selenization. *Appl. Surf. Sci.* **2021**, *546*, 149076, doi:10.1016/j.apsusc.2021.149076.

40. Chen, S.-Y.; Goldstein, T.; Venkataraman, D.; Ramasubramaniam, A.; Yan, J. Activation of New Raman Modes by Inversion Symmetry Breaking in Type II Weyl Semimetal Candidate T'-MoTe2. *Nano Lett.* **2016**, *16*, 5852–5860, doi:10.1021/acs.nanolett.6b02666.

41. Kang, S.; Won, D.; Yang, H.; Lin, C.-H.; Ku, C.-S.; Chiang, C.-Y.; Kim, S.; Cho, S. Phase-Controllable Laser Thinning in $MoTe_2$. *Appl. Surf. Sci.* **2021**, *563*, 150282, doi:10.1016/j.apsusc.2021.150282.

42. Chen, Z.; Nan, H.; Liu, Z.; Wang, X.; Gu, X.; Xiao, S. Effect of Thermal Conductivity of Substrate on Laser-induced Phase Transition of $MoTe_2$. *Journal of Raman Spectroscopy* **2019**, *50*, 755–761, doi:10.1002/jrs.5559.

43. Li, Z.; Zhan, F.; Ge, H.; Yan, F.; Tong, Q.; Luo, J.; Xie, S.; Wang, R.; Liu, Y.; Zhang, Q.; et al. The Critical Role of Interlayer Charge Transfer and Charge Redistribution Inducing the Formation

of Phase-Pure Monolayer 1T′-$MoTe_2$. *ACS Nano* **2025**, *19*, 16685–16695, doi:10.1021/acsnano.5c00944.

44. Qin, J.-K.; Liao, P.-Y.; Si, M.; Gao, S.; Qiu, G.; Jian, J.; Wang, Q.; Zhang, S.-Q.; Huang, S.; Charnas, A.; et al. Raman Response and Transport Properties of Tellurium Atomic Chains Encapsulated in Nanotubes. *Nat. Electron.* **2020**, *3*, 141–147, doi:10.1038/s41928-020-0365-4.

45. Khatun, S.; Banerjee, A.; Pal, A.J. Nonlayered Tellurene as an Elemental 2D Topological Insulator: Experimental Evidence from Scanning Tunneling Spectroscopy. *Nanoscale* **2019**, *11*, 3591–3598, doi:10.1039/C8NR09760G.

46. Silva, R.R.; Mejia, H.A.G.; Ribeiro, S.J.L.; Shrestha, L.K.; Ariga, K.; Oliveira Jr., O.N.; Camargo, V.R.; Maia, L.J.Q.; Araújo, C.B. Facile Synthesis of Tellurium Nanowires and Study of Their Third-Order Nonlinear Optical Properties. *J. Braz. Chem. Soc.* **2016**, doi:10.5935/0103-5053.20160145.

47. Li, G.; Cui, X.; Tan, C.; Lin, N. Solvothermal Synthesis of Polycrystalline Tellurium Nanoplates and Their Conversion into Single Crystalline Nanorods. *RSC Adv.* **2014**, *4*, 954–958, doi:10.1039/C3RA41801D.

48. Piskorski, M.; Lutsyk, I.; Ryś, W.; Ster, M. Le; Ogorzałek-Sory, Z.; Binder, J.; Toczek, K.; Nadolska, A.; Dunal, R.; Przybysz, P.; et al. The Integration of Raman Spectrometer with Glove Box for High-Purity Investigation in an Inert Gas Condition. *Measurement* **2025**, *251*, 117190, doi:10.1016/j.measurement.2025.117190.

49. Seredyński, B.; Bożek, R.; Suffczyński, J.; Piwowar, J.; Sadowski, J.; Pacuski, W. Molecular Beam Epitaxy Growth of MoTe2 on Hexagonal Boron Nitride. *J. Cryst. Growth* **2022**, *596*, 126806, doi:10.1016/j.jcrysgro.2022.126806.

50. Chen, B.; Sahin, H.; Suslu, A.; Ding, L.; Bertoni, M.I.; Peeters, F.M.; Tongay, S. Environmental Changes in MoTe2 Excitonic Dynamics by Defects-Activated Molecular Interaction. *ACS Nano* **2015**, *9*, 5326–5332, doi:10.1021/acsnano.5b00985.

51. Ogorzałek, Z.; Seredyński, B.; Kret, S.; Kwiatkowski, A.; Korona, K.P.; Grzeszczyk, M.; Mierzejewski, J.; Wasik, D.; Pacuski, W.; Sadowski, J.; et al. Charge Transport in MBE-Grown 2H-$MoTe_2$ Bilayers with Enhanced Stability Provided by an $AlO_x$ Capping Layer. *Nanoscale* **2020**, *12*, 16535–16542, doi:10.1039/D0NR03148H.

52. Pham, T.T.; Castelino, R.; Felten, A.; Sporken, R. Study of Surface Oxidation and Recovery of Clean MoTe2 Films. *Surfaces and Interfaces* **2022**, *28*, 101681, doi:10.1016/j.surfin.2021.101681.

53. Zhu, H.; Wang, Q.; Cheng, L.; Addou, R.; Kim, J.; Kim, M.J.; Wallace, R.M. Defects and Surface Structural Stability of MoTe2 Under Vacuum Annealing. *ACS Nano* **2017**, *11*, 11005–11014, doi:10.1021/acsnano.7b04984.

54. Åhlgren, E.H.; Markevich, A.; Scharinger, S.; Fickl, B.; Zagler, G.; Herterich, F.; McEvoy, N.; Mangler, C.; Kotakoski, J. Atomic-Scale Oxygen-Mediated Etching of 2D MoS2 and MoTe2. *Adv. Mater. Interfaces* **2022**, *9*, doi:10.1002/admi.202200987.

55. Polyanskiy, M.N. Refractiveindex.Info Database of Optical Constants. *Sci. Data* **2024**, *11*, 94, doi:10.1038/s41597-023-02898-2.

56. Liu, Z.; Peng, B.; Tsai, Y.-H.J.; Zhang, A.; Xu, M.; Zang, W.; Yan, X.; Xing, L.; Pan, X.; Duan, X.; et al. Pt Catalyst Protected by Graphene Nanopockets Enables Lifetimes of over 200,000 h for Heavy-Duty Fuel Cell Applications. *Nat. Nanotechnol.* **2025**, doi:10.1038/s41565-025-01895-3.

57. Le Ster, M.; Chan, J.R.; Ruck, B.J.; Brown, S.A.; Natali, F. Removable Capping Layer for Air-Sensitive GdN. *Nanotechnology* **2020**, *31*, 275709, doi:10.1088/1361-6528/ab82d3.

58. Ryś, W.; Lutsyk, I.; Le Ster, M.; Przybysz, P.; Sławińska, J.; Dąbrowski, P.; Rogala, M.; Szałowski, K.; Sobol, T.; Partyka-Jankowska, E.; et al. The Coexistence of Dirac Cones and Fermi Arcs in a Graphene/$WTe_2$ Heterostructure. *Nanoscale* **2025**, *17*, 26835–26844, doi:10.1039/D5NR03824C.

59. Arora, H.; Jung, Y.; Venanzi, T.; Watanabe, K.; Taniguchi, T.; Hübner, R.; Schneider, H.; Helm, M.; Hone, J.C.; Erbe, A. Effective Hexagonal Boron Nitride Passivation of Few-Layered InSe and GaSe to Enhance Their Electronic and Optical Properties. *ACS Appl. Mater. Interfaces* **2019**, *11*, 43480–43487, doi:10.1021/acsami.9b13442.

60. Rosati, R.; Paradisanos, I.; Huang, L.; Gan, Z.; George, A.; Watanabe, K.; Taniguchi, T.; Lombez, L.; Renucci, P.; Turchanin, A.; et al. Interface Engineering of Charge-Transfer Excitons in 2D Lateral Heterostructures. *Nat. Commun.* **2023**, *14*, 2438, doi:10.1038/s41467-023-37889-9.

61. Lee, H.S.; Sung, J.; Shin, D.-J.; Gong, S.-H. The Impact of HBN Layers on Guided Exciton–Polariton Modes in WS2 Multilayers. *Nanophotonics* **2024**, *13*, 1475–1482, doi:10.1515/nanoph-2023-0822.

62. Nutting, D.; Felix, J.F.; Tillotson, E.; Shin, D.-W.; Sanctis, A. De; Chang, H.; Cole, N.; Russo, S.; Woodgate, A.; Leontis, I.; et al. Heterostructures Formed through Abraded van Der Waals Materials. *Nat. Commun.* **2020**, *11*, 3047, doi:10.1038/s41467-020-16717-4.

63. Jin, C.; Kim, J.; Suh, J.; Shi, Z.; Chen, B.; Fan, X.; Kam, M.; Watanabe, K.; Taniguchi, T.; Tongay, S.; et al. Interlayer Electron–Phonon Coupling in WSe2/HBN Heterostructures. *Nat. Phys.* **2017**, *13*, 127–131, doi:10.1038/nphys3928.

64. Cianci, S.; Blundo, E.; Tuzi, F.; Pettinari, G.; Olkowska-Pucko, K.; Parmenopoulou, E.; Peeters, D.B.L.; Miriametro, A.; Taniguchi, T.; Watanabe, K.; et al. Spatially Controlled Single Photon Emitters in HBN-Capped WS2 Domes. *Adv. Opt. Mater.* **2023**, *11*, doi:10.1002/adom.202202953.

65. Ogawa, S.; Fukushima, S.; Shimatani, M. Hexagonal Boron Nitride for Photonic Device Applications: A Review. *Materials* **2023**, *16*, 2005, doi:10.3390/ma16052005.

66. Wlasny, I.; Dabrowski, P.; Rogala, M.; Kowalczyk, P.J.; Pasternak, I.; Strupinski, W.; Baranowski, J.M.; Klusek, Z. Role of Graphene Defects in Corrosion of Graphene-Coated Cu(111) Surface. *Appl. Phys. Lett.* **2013**, *102*, 111601, doi:10.1063/1.4795861.

67. Yang, X.; Zhang, R.; Pu, J.; He, Z.; Xiong, L. 2D Graphene and HBN Layers Application in Protective Coatings. *Corrosion Reviews* **2021**, *39*, 93–107, doi:10.1515/corrrev-2020-0080.

68. Truong, Q.D.; Hung, N.T.; Nakayasu, Y.; Nayuki, K.; Sasaki, Y.; Murukanahally Kempaiah, D.; Yin, L.-C.; Tomai, T.; Saito, R.; Honma, I. Inversion Domain Boundaries in $MoSe_2$ Layers. *RSC Adv.* **2018**, *8*, 33391–33397, doi:10.1039/C8RA07205A.

69. Ryś, W.; Lutsyk, I.; Szałowski, K.; Ster, M. Le; Rogala, M.; Piskorski, M.; Krukowski, P.; Dąbrowski, P.; Dunal, R.; Nadolska, A.; et al. Degradation of BiTeCl Induced by Thermal and Laser Treatment. *Sci. Rep.* **2025**, *15*, 15936, doi:10.1038/s41598-025-00907-5.

70. Zhang, Y.; Li, X.; Li, Y.; Wu, D.; Miao, X.; Li, L.; Min, T.; Pan, Y. Transport Property Evolution in 2H-MoTe $_{2-x}$ Mediated by Te-Deficiency-Induced Mirror Twin Boundary Networks. *Small Struct.* **2024**, *5*, doi:10.1002/sstr.202400027.

71. Yu, Y.; Wang, G.; Qin, S.; Wu, N.; Wang, Z.; He, K.; Zhang, X.-A. Molecular Beam Epitaxy Growth of Atomically Ultrathin $MoTe_2$ Lateral Heterophase Homojunctions on Graphene Substrates. *Carbon N. Y.* **2017**, *115*, 526–531, doi:10.1016/j.carbon.2017.01.026.

72. Dong, L.; Wang, G.-Y.; Zhu, Z.; Zhao, C.-X.; Yang, X.-Y.; Li, A.-M.; Chen, J.-L.; Guan, D.-D.; Li, Y.-Y.; Zheng, H.; et al. Charge Density Wave States in 2H-MoTe $_2$ Revealed by Scanning Tunneling Microscopy. *Chinese Physics Letters* **2018**, *35*, 066801, doi:10.1088/0256-307X/35/6/066801.

73. Sun, L.; Ding, M.; Li, J.; Yang, L.; Lou, X.; Xie, Z.; Zhang, W.; Chang, H. Phase-Controlled Large-Area Growth of $MoTe_2$ and $MoTe_{2-x}O_x$/$MoTe_2$ Heterostructures for Tunable Memristive Behavior. *Appl. Surf. Sci.* **2019**, *496*, 143687, doi:10.1016/j.apsusc.2019.143687.

74. Pace, S.; Martini, L.; Convertino, D.; Keum, D.H.; Forti, S.; Pezzini, S.; Fabbri, F.; Mišeikis, V.; Coletti, C. Synthesis of Large-Scale Monolayer 1T′-MoTe $_2$ and Its Stabilization *via* Scalable HBN Encapsulation. *ACS Nano* **2021**, *15*, 4213–4225, doi:10.1021/acsnano.0c05936.

75. Fraser, J.P.; Masaityte, L.; Zhang, J.; Laing, S.; Moreno-López, J.C.; McKenzie, A.F.; McGlynn, J.C.; Panchal, V.; Graham, D.; Kazakova, O.; et al. Selective Phase Growth and Precise-Layer Control in $MoTe_2$. *Commun. Mater.* **2020**, *1*, 48, doi:10.1038/s43246-020-00048-4.

76. Naylor, C.H.; Parkin, W.M.; Ping, J.; Gao, Z.; Zhou, Y.R.; Kim, Y.; Streller, F.; Carpick, R.W.; Rappe, A.M.; Drndić, M.; et al. Monolayer Single-Crystal 1T′-MoTe $_2$ Grown by Chemical Vapor Deposition Exhibits Weak Antilocalization Effect. *Nano Lett.* **2016**, *16*, 4297–4304, doi:10.1021/acs.nanolett.6b01342.

77. Pálinkás, A.; Kálvin, G.; Vancsó, P.; Kandrai, K.; Szendrő, M.; Németh, G.; Németh, M.; Pekker, Á.; Pap, J.S.; Petrik, P.; et al. The Composition and Structure of the Ubiquitous Hydrocarbon Contamination on van Der Waals Materials. *Nat. Commun.* **2022**, *13*, 6770, doi:10.1038/s41467-022-34641-7.

78. Onda, K.; Li, B.; Petek, H. Two-Photon Photoemission Spectroscopy of TiO2(110) Surfaces Modified by Defects and O2 or H2O Adsorbates. *Phys. Rev. B* **2004**, *70*, 045415, doi:10.1103/PhysRevB.70.045415.

79. Musumeci, F.; Pollack, G.H. Influence of Water on the Work Function of Certain Metals. *Chem. Phys. Lett.* **2012**, *536*, 65–67, doi:10.1016/j.cplett.2012.03.094.

80. Rietwyk, K.J.; Keller, D.A.; Ginsburg, A.; Barad, H.; Priel, M.; Majhi, K.; Yan, Z.; Tirosh, S.; Anderson, A.Y.; Ley, L.; et al. Universal Work Function of Metal Oxides Exposed to Air. *Adv. Mater. Interfaces* **2019**, *6*, doi:10.1002/admi.201802058.

81. Mishra, S.; Kabiraj, D.; Roy, A.; Ghosh, S. Effect of High-energy Light-ion Irradiation on SI-GaAs and GaAs:Cr as Observed by Raman Spectroscopy. *Journal of Raman Spectroscopy* **2012**, *43*, 344–350, doi:10.1002/jrs.3039.

82. Seredin, P. V.; Goloshchapov, D.L.; Arsentyev, I.N.; Nikolaev, D.N.; Pikhtin, N.A.; Slipchenko, S.O. Spectroscopic Studies of Integrated GaAs/Si Heterostructures. *Semiconductors* **2021**, *55*, 44–50, doi:10.1134/S1063782621010139.

83. Das, S.K.; Das, T.D.; Dhar, S. Properties of GaAsN Layers Grown from Melt Containing $Li_3$ N as Flux for Enhancing Nitrogen Dissolution. *Semicond. Sci. Technol.* **2011**, *26*, 085028, doi:10.1088/0268-1242/26/8/085028.

84. Rani, A.; DiCamillo, K.; Krylyuk, S.; Davydov, A. V.; Debnath, R.; Taheri, P.; Korman, C.E.; Paranjape, M.; Zaghloul, M.E. Control of Polarity in Multilayer MoTe2 Field-Effect Transistors by Channel Thickness. In Proceedings of the Low-Dimensional Materials and Devices 2018; Kobayashi, N.P., Talin, A.A., Davydov, A. V., Islam, M.S., Eds.; SPIE, September 11 2018; p. 41.

85. Arora, A.; Koperski, M.; Slobodeniuk, A.; Nogajewski, K.; Schmidt, R.; Schneider, R.; Molas, M.R.; de Vasconcellos, S.M.; Bratschitsch, R.; Potemski, M. Zeeman Spectroscopy of Excitons and Hybridization of Electronic States in Few-Layer $WSe_2$ , $MoSe_2$ and $MoTe_2$. *2d Mater.* **2018**, *6*, 015010, doi:10.1088/2053-1583/aae7e5.

86. Song, Q.; Wang, H.; Pan, X.; Xu, X.; Wang, Y.; Li, Y.; Song, F.; Wan, X.; Ye, Y.; Dai, L. Anomalous In-Plane Anisotropic Raman Response of Monoclinic Semimetal 1T'-MoTe2. *Sci. Rep.* **2017**, *7*, 1758, doi:10.1038/s41598-017-01874-2.

87. Byrnes, S.J. Multilayer Optical Calculations. *arXiv:1603.02720v5 [physics.comp-ph]* 2020.

88. Werner, W.S.M.; Glantschnig, K.; Ambrosch-Draxl, C. Optical Constants and Inelastic Electron-Scattering Data for 17 Elemental Metals. *J. Phys. Chem. Ref. Data* **2009**, *38*, 1013–1092, doi:10.1063/1.3243762.

89. Szałowski, K. Janus Monolayer of 1T-TaSSe: A Computational Study. *Materials* **2024**, *17*, 4591, doi:10.3390/ma17184591.

90. Samanta, M.; Ghosh, T.; Chandra, S.; Biswas, K. Layered Materials with 2D Connectivity for Thermoelectric Energy Conversion. *J. Mater. Chem. A Mater.* **2020**, *8*, 12226–12261, doi:10.1039/D0TA00240B.

91. Jana, M.K.; Singh, A.; Sampath, A.; Rao, C.N.R.; Waghmare, U. V. Structure and Electron-Transport Properties of Anion-Deficient $MoTe_2$ : A Combined Experimental and Theoretical Study. *Z. Anorg. Allg. Chem.* **2016**, *642*, 1386–1396, doi:10.1002/zaac.201600314.

92. Perdew, J.P. Density Functional Theory and the Band Gap Problem. *Int. J. Quantum Chem.* **2009**, *28*, 497–523, doi:10.1002/qua.560280846.

93. Bae, S.; Sugiyama, N.; Matsuo, T.; Raebiger, H.; Shudo, K.; Ohno, K. Defect-Induced Vibration Modes of Ar+ -Irradiated $MoS_2$. *Phys. Rev. Appl.* **2017**, *7*, 024001, doi:10.1103/PhysRevApplied.7.024001.

94. Huang, T.-X.; Cong, X.; Wu, S.-S.; Lin, K.-Q.; Yao, X.; He, Y.-H.; Wu, J.-B.; Bao, Y.-F.; Huang, S.-C.; Wang, X.; et al. Probing the Edge-Related Properties of Atomically Thin $MoS_2$ at Nanoscale. *Nat. Commun.* **2019**, *10*, 5544, doi:10.1038/s41467-019-13486-7.

95. Niwase, K. Raman Spectroscopy for Quantitative Analysis of Point Defects and Defect Clusters in Irradiated Graphite. *Int. J. Spectrosc.* **2012**, *2012*, 1–14, doi:10.1155/2012/197609.

96. O'Leary, W.; Grumet, M.; Kaiser, W.; Bučko, T.; Rupp, J.L.M.; Egger, D.A. Rapid Characterization of Point Defects in Solid-State Ion Conductors Using Raman Spectroscopy, Machine-Learning Force Fields, and Atomic Raman Tensors. *J. Am. Chem. Soc.* **2024**, *146*, 26863–26876, doi:10.1021/jacs.4c07812.

97. Shi, J.; Bie, Y.-Q.; Zong, A.; Fang, S.; Chen, W.; Han, J.; Cao, Z.; Zhang, Y.; Taniguchi, T.; Watanabe, K.; et al. Intrinsic 1T′ Phase Induced in Atomically Thin 2H-MoTe2 by a Single Terahertz Pulse. *Nat. Commun.* **2023**, *14*, 5905, doi:10.1038/s41467-023-41291-w.

98. Awate, S.S.; Xu, K.; Liang, J.; Katz, B.; Muzzio, R.; Crespi, V.H.; Katoch, J.; Fullerton-Shirey, S.K. Strain-Induced 2H to 1T′ Phase Transition in Suspended $MoTe_2$ Using Electric Double Layer Gating. *ACS Nano* **2023**, *17*, 22388–22398, doi:10.1021/acsnano.3c04701.

99. Tan, Y.; Luo, F.; Zhu, M.; Xu, X.; Ye, Y.; Li, B.; Wang, G.; Luo, W.; Zheng, X.; Wu, N.; et al. Controllable 2H-to-1T′ Phase Transition in Few-Layer $MoTe_2$. *Nanoscale* **2018**, *10*, 19964–19971, doi:10.1039/C8NR06115G.

100. Mattausch, A.; Pankratov, O. Ab Initio Study of Graphene on SiC. *Phys. Rev. Lett.* **2007**, *99*, 076802, doi:10.1103/PhysRevLett.99.076802.

101. Emtsev, K. V.; Speck, F.; Seyller, Th.; Ley, L.; Riley, J.D. Interaction, Growth, and Ordering of Epitaxial Graphene on SiC{0001} Surfaces: A Comparative Photoelectron Spectroscopy Study. *Phys. Rev. B* **2008**, *77*, 155303, doi:10.1103/PhysRevB.77.155303.

102. Pradeepkumar, A.; Amjadipour, M.; Mishra, N.; Liu, C.; Fuhrer, M.S.; Bendavid, A.; Isa, F.; Zielinski, M.; Sirikumara, H.I.; Jayasekara, T.; et al. P-Type Epitaxial Graphene on Cubic Silicon Carbide on Silicon for Integrated Silicon Technologies. *ACS Appl. Nano Mater.* **2020**, *3*, 830–841, doi:10.1021/acsanm.9b02349.

103. Seredyński, B.; Ogorzałek, Z.; Zajkowska, W.; Bożek, R.; Tokarczyk, M.; Suffczyński, J.; Kret, S.; Sadowski, J.; Gryglas-Borysiewicz, M.; Pacuski, W. Molecular Beam Epitaxy of a 2D Material Nearly Lattice Matched to a 3D Substrate: $NiTe_2$ on GaAs. *Cryst. Growth Des.* **2021**, *21*, 5773–5779, doi:10.1021/acs.cgd.1c00673.

104. Siegfried, P.E.; Bhandari, H.; Qi, J.; Ghimire, R.; Joshi, J.; Messegee, Z.T.; Beeson, W.B.; Liu, K.; Ghimire, M.P.; Dang, Y.; et al. $CoTe_2$ : A Quantum Critical Dirac Metal with Strong Spin Fluctuations. *Advanced Materials* **2023**, *35*, doi:10.1002/adma.202300640.

105. Ashraf, M.W.; Haider, S.I.; Solangi, A.R.; Memon, A.F. Toxicity of Tellurium and Its Compounds. *Physical Sciences Reviews* **2023**, *8*, 4375–4390, doi:10.1515/psr-2021-0112.

106. Genchi, G.; Carocci, A.; Lauria, G.; Sinicropi, M.S.; Catalano, A. Nickel: Human Health and Environmental Toxicology. *Int. J. Environ. Res. Public Health* **2020**, *17*, 679, doi:10.3390/ijerph17030679.

107. Leyssens, L.; Vinck, B.; Van Der Straeten, C.; Wuyts, F.; Maes, L. Cobalt Toxicity in Humans—A Review of the Potential Sources and Systemic Health Effects. *Toxicology* **2017**, *387*, 43–56, doi:10.1016/j.tox.2017.05.015.

108. Nečas, D.; Klapetek, P. Gwyddion: An Open-Source Software for SPM Data Analysis. *Open Physics* **2012**, *10*, 181–188, doi:10.2478/s11534-011-0096-2.

109. Horcas, I.; Fernandez, R.; Gomez-Rodriguez, J.M.; Colchero, J.; Gomez-Herrero, J.; Baro, A.M. {WSXM}: A Software for Scanning Probe Microscopy and a Tool for Nanotechnology. *Rev. Sci. Instrum.* **2007**, *78*, 13705, doi:10.1063/1.2432410.

110. Piskorski, M.; Rogala, M.; Dąbrowski, P.; Lutsyk, I.; Kozłowski, W.; Le Ster, M.; Kowalczyk, P.J.; Radecki, Ł.; Krukowski, P. High-Precision Spatial Mapping Control for a Glove Box-Integrated Raman Spectrometer. *Opto-Electronics Review* **2025**, 155901–155901, doi:10.24425/opelre.2025.155901.

111. Radecki, Ł.; Piskorski, M.; Kowalczyk, P.J.; Krukowski, P. Sample Positioning Control System Using Inductive Sensors in a Mapping Raman Spectroscopy. *PRZEGLĄD ELEKTROTECHNICZNY* **2026**, *1*, 24–30, doi:10.15199/48.2026.04.4.

112. Piskorski, M.; Krukowski, P.; Kozłowski, W.; Rogala, M.; Dąbrowski, P.; Lutsyk, I.; Kowalczyk, D.A.; Le Ster, M.; Sałagan, K.; Nadolska, A.; et al. The Vibration Registration System with the Use of a Seismic Sensor and a Real-Time Spectrum Analyzer in the Room Intended for the TERS-STM System Installations. *Przegląd Elektrotechniczny* **2023**, *1*, 286–289, doi:10.15199/48.2023.11.55.

113. Schindelin, J.; Arganda-Carreras, I.; Frise, E.; Kaynig, V.; Longair, M.; Pietzsch, T.; Preibisch, S.; Rueden, C.; Saalfeld, S.; Schmid, B.; et al. Fiji: An Open-Source Platform for Biological-Image Analysis. *Nat. Methods* **2012**, *9*, 676–682, doi:10.1038/nmeth.2019.

114. Giannozzi, P.; Baroni, S.; Bonini, N.; Calandra, M.; Car, R.; Cavazzoni, C.; Ceresoli, D.; Chiarotti, G.L.; Cococcioni, M.; Dabo, I.; et al. QUANTUM ESPRESSO: A Modular and Open-Source Software Project for Quantum Simulations of Materials. *Journal of Physics: Condensed Matter* **2009**, *21*, 395502, doi:10.1088/0953-8984/21/39/395502.

115. Giannozzi, P.; Andreussi, O.; Brumme, T.; Bunau, O.; Buongiorno Nardelli, M.; Calandra, M.; Car, R.; Cavazzoni, C.; Ceresoli, D.; Cococcioni, M.; et al. Advanced Capabilities for Materials Modelling with Quantum ESPRESSO. *Journal of Physics: Condensed Matter* **2017**, *29*, 465901, doi:10.1088/1361-648X/aa8f79.

116. Hohenberg, P.; Kohn, W. Inhomogeneous Electron Gas. *Physical Review* **1964**, *136*, B864–B871, doi:10.1103/PhysRev.136.B864.

117. Bengtsson, L. Dipole Correction for Surface Supercell Calculations. *Phys. Rev. B* **1999**, *59*, 12301–12304, doi:10.1103/PhysRevB.59.12301.

118. Perdew, J.P.; Ruzsinszky, A.; Csonka, G.I.; Vydrov, O.A.; Scuseria, G.E.; Constantin, L.A.; Zhou, X.; Burke, K. Restoring the Density-Gradient Expansion for Exchange in Solids and Surfaces. *Phys. Rev. Lett.* **2008**, *100*, 136406, doi:10.1103/PhysRevLett.100.136406.

119. Kresse, G.; Joubert, D. From Ultrasoft Pseudopotentials to the Projector Augmented-Wave Method. *Phys. Rev. B* **1999**, *59*, 1758–1775, doi:10.1103/PhysRevB.59.1758.

120. Dal Corso, A. Pseudopotentials Periodic Table: From H to Pu. *Comput. Mater. Sci.* **2014**, *95*, 337–350, doi:10.1016/j.commatsci.2014.07.043.

121. Grimme, S.; Antony, J.; Ehrlich, S.; Krieg, H. A Consistent and Accurate *Ab Initio* Parametrization of Density Functional Dispersion Correction (DFT-D) for the 94 Elements H-Pu. *J. Chem. Phys.* **2010**, *132*, 154104, doi:10.1063/1.3382344.

122. Methfessel, M.; Paxton, A.T. High-Precision Sampling for Brillouin-Zone Integration in Metals. *Phys. Rev. B* **1989**, *40*, 3616–3621, doi:10.1103/PhysRevB.40.3616.

123. Blöchl, P.E.; Jepsen, O.; Andersen, O.K. Improved Tetrahedron Method for Brillouin-Zone Integrations. *Phys. Rev. B* **1994**, *49*, 16223–16233, doi:10.1103/PhysRevB.49.16223.

**Supplementary Information**

# Thickness-dependent degradation and optical access in epitaxial 2H-$MoTe_2$ protected by metallic capping layers

Wojciech Ryś[1], Iaroslav Lutsyk[1], Michał Piskorski[1], Maxime Le Ster[1], Maciej Rogala[1], Paweł Dąbrowski[1], Paweł Krukowski[1], Katarzyna Ranoszek-Soliwoda[2], Jarosław Grobelny[2], Zuzanna Ogorzałek-Sory[3], Wojciech Pacuski[3], Janusz Sadowski[4,5], Marta Gryglas-Borysiewicz[3], Karol Szałowski[1], Paweł J. Kowalczyk[1*]

[1] University of Lodz, Faculty of Physics and Applied Informatics, Pomorska 149/153, 90-236 Łódź, Poland

[2] University of Lodz, Faculty of Chemistry, Department of Materials Technology and Chemistry, Pomorska 163, 90-236 Lodz, Poland

[3] University of Warsaw, Faculty of Physics, Pasteura 5, 02-093 Warsaw, Poland

[4] Polish Academy of Sciences, Institute of Physics, Aleja Lotnikow 32/46, Warsaw, Poland

[5] Ensemble3 Centre of Excellence, Wolczynska 133, Warsaw, Poland

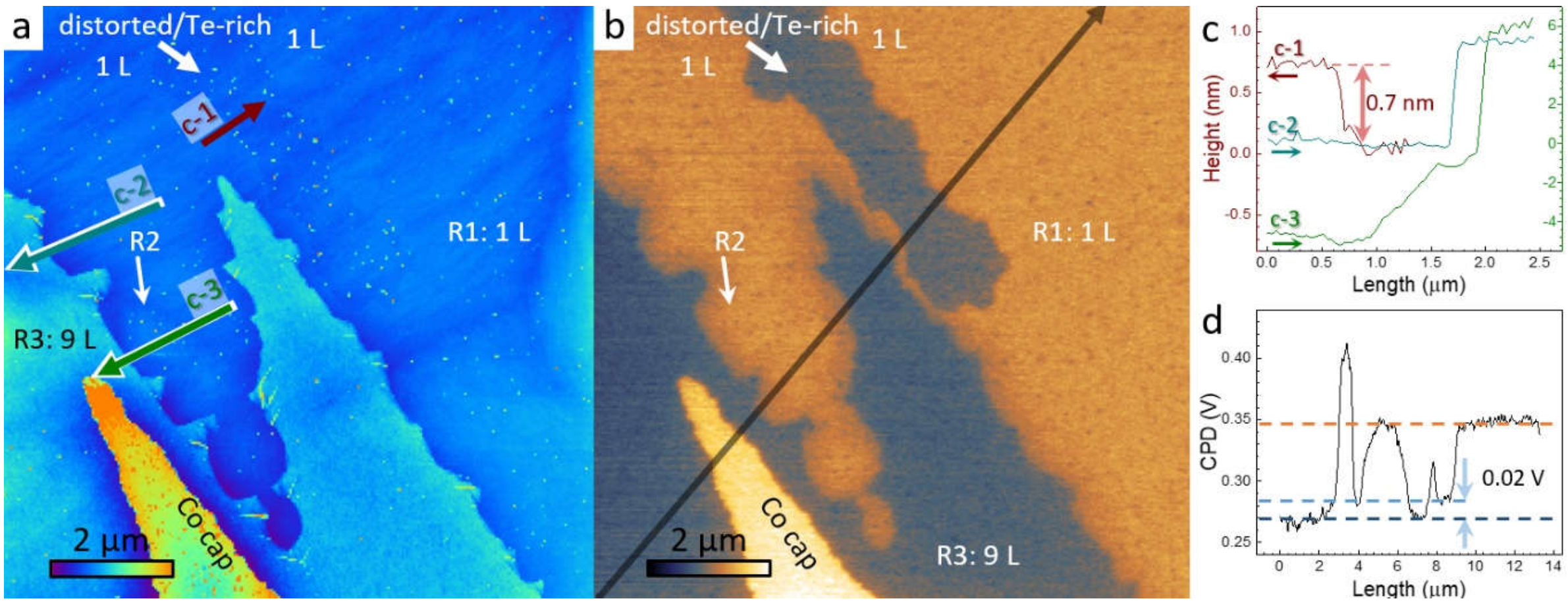

*SI Fig. S1 (a) AFM image showing the presence of a distorted region (indicated using white arrow). (b) Corresponding CPD image recorded in KPM mode. (c) Three cross-section plots denoted c-1, c-2 and c-3 along arrows shown in (a). The profile c-1 shows terrace height at the level of 0.7 nm, which corresponds to 1 L drop from the base height of 1 L. In turn, profile c-2 shows that the height of the R3 region is approx. 5.5 nm, which corresponds to 8 L measured from the base of a 1 L thick region. This indicates that the R3 region is approx. 9 L thick under the assumption of 1 L thickness of the base layer. Profile c-3 is extracted from the region in which an anomalous Raman signal was measured. This region is indicated using a white arrow in (a). Both profile c-3 and the indicated region in (a) reveal the presence of grain, which can be responsible for the increase of Raman signal at 124 cm$^{-1}$ shown in Fig. 3. (d) Profile showing variation of CPD along the line shown in (b).*

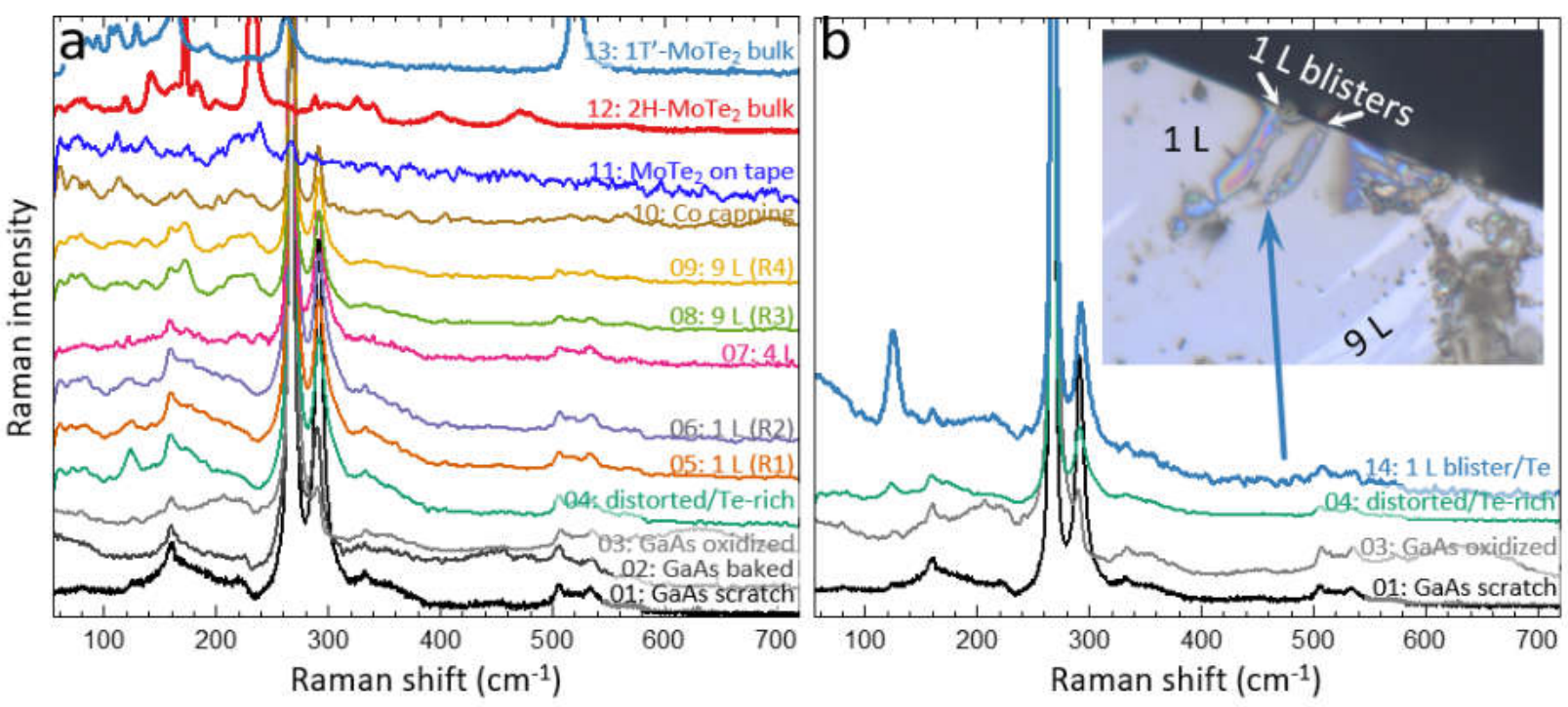

*SI Fig. S2. (a) Full Range Raman spectra recorded using 1800 l/mm grating including 01: scratched region, 02: UHV-baked GaAs, 03: the oxidized back side of the crystal, and 04: distorted/Te-rich (see marked region in SI Fig. S1). Spectra recorded on $MoTe_2$ films of different thicknesses are also shown, including 05: R1 and 06: R2 1 L thick regions, 07: 4 L $MoTe_2$ exfoliated and measured in an Ar protective atmosphere, 08: R3 and 09: R4 9 L thick regions, and 10: $MoTe_2$ covered by a capping layer, as well as 11: $MoTe_2$ exfoliated on adhesive tape. Reference spectra recorded on 12: 2H-$MoTe_2$ and 13: 1T'-$MoTe_2$ bulk crystals. (b) Comparison of the distorted $MoTe_2$ layer characterized by the Te mode at approx. 124 $cm^{-1}$ (spectrum 04) with spectra recorded on 1 L $MoTe_2$ blister locally characterized by strong Te modes (spectrum 14). Substrate Raman spectra of 01: GaAs scratch and 03: oxidized backside of GaAs are added as reference. The inset shows an optical microscopy image of the sample edge with several blisters formed after exfoliation attempts. The arrow indicates the location in which a very strong Te-related signal was recorded.*

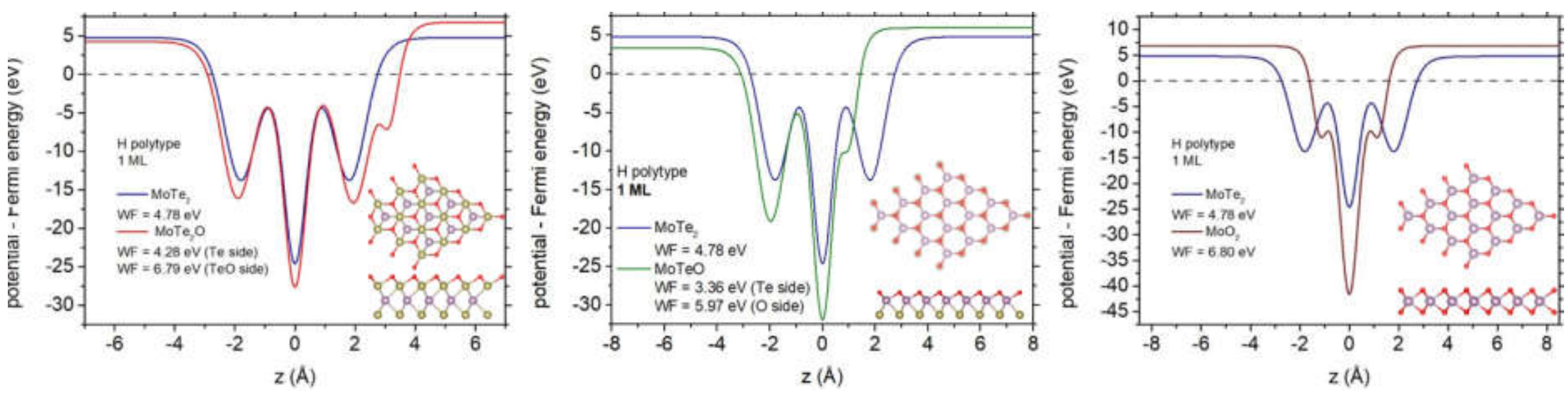

*SI Fig. S3. DFT calculated work function i.e., difference of potential and Fermi energy as a function of distance for single layer oxides of H polytype $MoTe_2$. The Z scale in plots was shifted in a way that Z=0 corresponds to the location of the Mo atom in the unit cell. Models for each oxide are shown as insets.*

**SI Tab. S1.** Experiment and sample history.

| Dataset / figure | Nominal thickness | Cap | Treatment | Age / exposure | Technique |
|---|---|---|---|---|---|
| **(1) Main samples** | 10 L | ~20 nm Co | | | |
| **Fig.1 e,f; XPS 01** | | Partially Co | Decapped in UHV | Fresh | XPS |
| **Fig. 2d; CPD** | | Partially Co | Air | 1-7 h | KPM/AFM |
| **Fig.1 e,f; XPS 02** | | Partially Co | Air | 7 h | XPS |
| **Fig.1 e,f; XPS 03** | | Partially Co | Air | ~2.5y | XPS |
| **Fig. 2a-c, Fig. 5 & Fig. S1; morphology, CPD** | | Partially Co | Air | ~2.5y | OM/AFM/KPM |
| **Fig. 3a-c; Raman spectrum 04-06, 08-10; Raman maps** | | Partially Co | Air | ~2.5y | Raman |
| **Fig. 3a; Raman spectrum 02 (GaAs)** | | - | Backside, baking in UHV at ~700 °C for 20 min | ~2.5y | Raman |
| **Fig. S2b; Raman spectrum 14 (blister)** | | Partially Co | Old sample broken into pieces in Ar | ~4.5y / fresh | Raman |
| **(2) Ar sample** | 4 L | ~40 nm Ni | | | |
| **Fig.1 d; STM** | | Partially Ni | Decapped in UHV | Fresh | STM |
| **Fig. 3a; Raman spectrum 07** | | Partially Ni | Decapped in Ar | Fresh | Raman |
| **(3) Peeled off** | 4 L | ~120 nm Ni | | | |
| **Fig. 3a; Raman spectrum 11** | | Ni | Peeled off in Ar using tape, tape with $MoTe_2$ is measured | Fresh | Raman |
| **(4) GaAs reference never exposed to Te** | | | | | |
| **Fig. 3a; Raman spectrum 01** | | | Scratched in Ar | Fresh | Raman |
| **Fig. 3a; Raman spectrum 03** | | | backside | A few years old, exposed to the air | Raman |
| **(5) 2H-$MoTe_2$ reference** | | | | | |
| **Fig. 3a; Raman spectrum 12** | | | exfoliated in Ar | Fresh | Raman |
| **(6) 1T'-$MoTe_2$ reference** | | | | | |
| **Fig. 3a; Raman spectrum 13** | | | exfoliated in Ar | Fresh | Raman |

**SI Tab. S2.** XPS fitting parameters used for Mo 3d and Te 3d spectra shown in Fig. 1e and 1f.

| Region | Component | Assignment | BE 3d5/2 (eV) | BE 3d3/2 (eV) | FWHM 3d5/2 (eV) | FWHM 3d3/2 (eV) |
|---|---|---|---|---|---|---|
| **01**. UHV exfoliated $MoTe_2$ sample | | | | | | |
| Mo 3d | Metallic | Mo-Te | 227.9 | 231.0 | 1.17 | 1.23 |
| Mo 3d | Oxide | Mo-O | - | - | - | |
| Te 3d | Metallic | Te-Mo | 573.0 | 583.3 | 1.40 | 1.36 |
| Te 3d | Oxide | Te-O | - | - | - | |
| **02**. $MoTe_2$ exposed to air for 7 h | | | | | | |
| Mo 3d | Metallic | Mo-Te | 228.0 | 231.1 | 1.27 | 1.42 |
| Mo 3d | Oxide | Mo-O | 232.4 | 235.5 | 1.80 | 1.81 |
| Te 3d | Metallic | Te-Mo | 573.0 | 583.4 | 1.42 | 1.42 |
| Te 3d | Oxide | Te-O | 576.4 | 586.8 | 1.64 | 1.63 |
| **03**. $MoTe_2$ exposed to air for 2.5 y | | | | | | |
| Mo 3d | Metallic | Mo-Te | 227.9 | 231.1 | 1.23 | 1.29 |
| Mo 3d | Oxide | Mo-O | 232.4 | 235.5 | 1.78 | 1.79 |
| Te 3d | Metallic | Te-Mo | 573.0 | 583.3 | 1.48 | 1.46 |
| Te 3d | Oxide | Te-O | 576.4 | 586.8 | 1.71 | 1.67 |